\pdfoutput=1
\documentclass[nofootinbib]{revtex4}
\usepackage{amssymb,amsmath,graphicx,color}
\usepackage{enumerate}
\usepackage{slashed}
\usepackage{hyperref}
\usepackage{xcolor}
\usepackage{verbatim}
\usepackage{caption} 
\usepackage[export]{adjustbox}

\hypersetup{
	colorlinks=true,
	linkcolor=blue,
	filecolor=blue,      
	urlcolor=blue, 
	bookmarks=true,
	citecolor=blue
}

\usepackage[paperwidth=230mm,paperheight=297mm,centering,hmargin=2cm,vmargin=2.5cm]{geometry}

\begin{document} 
\title{Operator Method and Recursion Relations for Inflationary Correlators} 
\author{Shek Kit Chu$^{1,2}$, Yi Wang$^{1,2}$, Siyi Zhou$^{1,2}$}  
\email{skchuab@connect.ust.hk, phyw@ust.hk, szhouah@connect.ust.hk}

\affiliation{${}^1$Department of Physics, The Hong Kong University of Science and Technology, \\
	Clear Water Bay, Kowloon, Hong Kong, P.R.China}
\affiliation{${}^2$Jockey Club Institute for Advanced Study, The Hong Kong University of Science and Technology, \\
	Clear Water Bay, Kowloon, Hong Kong, P.R.China} 

\begin{abstract}
We develop the systematics for applying operators on Minkowski correlation functions to get the inflationary correlation functions. Simple structures and recursion relations are known for Minkowski correlation functions. Using the operator technique, various novel recursion relations for inflationary correlation functions are obtained.
\end{abstract}

\maketitle

\section{Introduction}\label{Intro}
Despite the fact that the departures from a Gaussian probability distribution in the cosmic microwave background (CMB) and large scale structure (LSS) are small, primordial non-Gaussianities can play an important role in understanding the physics of inflation. From various shapes of non-Gaussianities, we can extract information about additional massive field contents during inflation \cite{Chen:2009we,Chen:2009zp,Baumann:2011nk,Noumi:2012vr,Arkani-Hamed:2015bza}, the initial conditions for inflation \cite{Jiang:2015hfa,Jiang:2016nok,Brandenberger:2016uzh}, the deviation from a standard kinetic term \cite{Chen:2006nt} and the shape of the inflaton potential during inflation \cite{Chen:2018brw,Chen:2018uul}. The simplest slow-roll inflation model predicts a non-vanishing non-Gaussianity which is suppressed by the slow-roll parameter~\cite{Maldacena:2002vr}. However,  general single field inflation~\cite{Chen:2006nt} can give much larger three-point correlation functions, and even higher-point correlation functions~\cite{Chen:2009bc,Lin:2010ua}. Thus studying the higher-point correlation functions and the relations between them have become an important topic in the coming era of precision cosmology.

The basic technique for computing the correlation functions is the in-in formalism~\cite{Weinberg:2005vy} (see also~\cite{Chen:2010xka,Wang:2013eqj} for reviews). The late time interacting vacuum states are evolved from an initial vacuum state using the evolution operator in the interaction picture. Equivalently, the formalism can be rewritten as the Schwinger-Keldysh formalism originally developed in~\cite{Schwinger:1960qe,Keldysh:1964ud,Feynman:1963fq} and later used in various aspects of physics~\cite{Landau10,Chou:1984es,Jordan:1986ug,Haehl:2016pec} and cosmology~\cite{Tsamis:1996qq,Tsamis:1996qm,Seery:2007we,vanderMeulen:2007ah,Seery:2008ax,Leblond:2010yq,Chen:2016nrs,Calzetta:1986ey,Prokopec:2010be,Gong:2016qpq}. See \cite{Chen:2017ryl} for a self-contained introduction of the Schwinger-Keldysh formalism. 

The computation of inflationary correlation functions are in general more tedious than the Minkowski correlation functions. There are two complications: the first one is that the time-dependence of the inflationary background makes the solution of the Klein-Gordon equation of the primordial curvature perturbation more involved than that of Minkowski. The other complication is due to the difference between time ordering, anti-time ordering and mixtures of them. In the Schwinger-Keldysh formalism, there are four types of propagators and we have to add up the contributions from all of them. For a diagram with $n$ vertices, there are in total $2^n$ contributions due to these different propagators. 

The first complication about more complicated mode function can be made simpler by extracting the time dependence of the inflationary background. It turns out that for any diagram of a correlation function in the de Sitter can be written as an differential operator acting on a diagram of a correlation function in the Minkowski space, both evaluated at $\tau = 0$. This is inspired by the discussion in \cite{Arkani-Hamed:2015bza} that in de Sitter space, by acting a particular differential operator on a three-point function of a conformally coupled scalar, we obtain the three-point function for massless field. This type of method is widely used to relate correlation functions of different types, for example, people apply the operator technique to find a relation between correlation functions in the presence of different types of massive field (different masses and spins) in AdS \cite{Isono:2018rrb} or dS \cite{Arkani-Hamed:2018kmz}. Similarly, in~\cite{Chu:2018ovy}, relations between Minkowski correlation functions and de Sitter correlation functions are proposed. 

In this article we improve the results of \cite{Chu:2018ovy} in a few aspects. In~\cite{Chu:2018ovy}, for the time integrals with different time orderings, we were unable to extract a unique operator (except for a particular form of interaction). In this work, we use a trick of ``deforming the internal momenta" and find that for a particular diagram ${\cal D}$, regardless of the time orderings, we can write
\begin{equation}\label{decomposition}
\langle \zeta_{{\bf k}_1} \cdots \zeta_{{\bf k}_n} \rangle'_{\cal D} = {\cal O}_{\cal D} \langle \phi_{{\bf k}_1} \cdots \phi_{{\bf k}_n} \rangle'_{\cal D} ~,
\end{equation}
where $\zeta$ and $\phi$ are massless scalars in de Sitter space and in Minkowski space, respectively, and ${\cal O}_{\cal D}$ is a differential operator associated to this diagram. In this expression, the internal momenta are deformed without being limited by momentum conservation. The deformation of the internal momentum looks constraining. However, as we will see later, the relation is powerful enough for deriving recursion relations, and one can take the momentum conserving limit after the whole procedure.

We develope a more systematic way of extracting the operator ${\cal O}_{\cal D}$ compared to the previous work. Other than getting rid of the complication from the time dependence due to the inflationary background, the more important implication of the decomposition in \eqref{decomposition} is that we can make use of the properties of the flat space correlation functions. In principle, for any relations for the flat space correlation function of the form
\begin{equation}
\langle \phi_{{\bf k}_1} \cdots \phi_{{\bf k}_n} \rangle'_{\cal D} = f(\{{\bf k}_i\},\{{\bf l}_j\}) ~,
\end{equation}
we can directly translate it into a relation for the corresponding inflationary correlation function by applying an appropriate operator:
\begin{equation}
{\cal O}_{\cal D} \langle \phi_{{\bf k}_1} \cdots \phi_{{\bf k}_n} \rangle'_{\cal D} = \langle \zeta_{{\bf k}_1} \cdots \zeta_{{\bf k}_n} \rangle'_{\cal D} = {\cal O}_{\cal D} f(\{{\bf k}_i\},\{{\bf l}_j\}) ~,
\end{equation}
where $\{{\bf k}_i\}$ is the set of all momenta of the external legs of ${\cal D}$ and $\{{\bf l}_j\}$ is the set of all momenta of the internal legs. One of the possible application is recursion relations. Even in the flat space, the computation gets more complicated when it goes to higher-points. Several useful recursion relations are developed in the context of scattering amplitudes. For example, the BCFW recursion relation~\cite{Britto:2005fq} makes use of the simplicity of on-shell amplitudes and simplify the higher-point scattering amplitudes into lower ones by putting the internal propagators on-shell. Lots of efforts has been made to derive analogous recursion relations in the curved space. For instance, in~\cite{Raju:2012zr,Raju:2012zs}, several recursion relations in $\rm AdS_4$ are derived using ideas very similar to BCFW. In \cite{Chu:2018ovy}, a recursion relation was derived by using recursion relations of the wave functions in the flat space \cite{Arkani-Hamed:2017fdk} (quite remarkably, these recursion relations posses some geometrical interpretations and some deep physics concepts like Lorentz invariance and unitarity can automatically emerge from this picture \cite{Arkani-Hamed:2018bjr, Benincasa:2018ssx}.) and some relations between flat space correlation function and flat space wave function, finally a recursion relation in inflationary correlation functions were derived. The operator method developed in this work can serve as a more direct way for deriving recursion relations for inflationary correlation functions in de Sitter space from recursion relations for flat space correlation functions, which applies for general form of interactions. We first derived a recursion relation for flat space correlation functions following a similar derivation for recursion relations of the wave function in the flat space in \cite{Arkani-Hamed:2017fdk}. Applying appropriate operators on these relations we finally obtain the desired recursion relations in de Sitter space.

The article will be organized as follows: In Section.~\ref{SK}, we review the the Schwinger-Keldysh diagrammatic rules for in-in correlation function calculation. In Section. \ref{operator}, we demonstrate the construction of the operator that transform the correlation function in the Minkowski space into de Sitter space. In Section. \ref{Rec}, we derive a recursion relation for tree diagrams in the massless scalar $\phi^3$ theory in the Minkowski space. Then we will use the operators constructed in Section. \ref{RecDS} to obtain a recursion relation in de Sitter space for a massless scalar field with cubic interactions.

\section{Schwinger-Keldysh Diagrammatic Rules}\label{SK}
In this section, we review the Schwinger-Keldysh diagramatic rules based on \cite{Chen:2017ryl} and setup the conventions for the later sections. Since we are considering the correlation function in both the Minkowski background and the inflationary background. It is convenient to introduce it in a general FRW background
\begin{align}
	ds^2 = a^2(\tau) (-d\tau^2+d\mathbf x^2)~,
\end{align}
where $\tau$ is the conformal time and $a(\tau)$ is the scale factor which equals to $1$ for the Minkowski background and is $-1/(H\tau)$ for the inflationary background, where $H\equiv\dot a/a$ is the Hubble parameter (we will use dots to denote derivative with respect to $t$ and prime for $\tau$). The conformal time is defined through $d\tau^2 = a^2(t) dt^2$, so in the Minkowski background, the conformal time is equal to the physical time $t$, which ranges from $-\infty$ to $\infty$. And in the inflationary background, $\tau$ ranges from $-\infty$ to $0$. 
Suppose we have a field theory defined by the action \footnote{Here we consider scalar field theory in the inflationary background. We remind the readers that for alternative background and also for tensor perturbations \cite{Li:2018wkt}, it is possible to apply similar techniques.}
\begin{align}
	S = \int d\tau d^3\mathbf x \, \mathcal L[\phi, \tau]~, 
\end{align}
where $\phi$ represents the field fluctuations, and $\mathcal L[\phi, \tau]$ is the Lagrangian of the theory starting from quadratic order in the fluctuations. The canonical conjugate momentum $\pi$ and the Hamiltonian are defined as
\begin{align}
	\pi \equiv \frac{\partial \mathcal L[\phi]}{\partial \phi'}, \quad \mathcal H[\pi,\phi] \equiv \pi \phi' - \mathcal L [\phi]~,
\end{align}
respectively.

In the canonical in-in formalism, the expectation value of the field fluctuations are defined through
\begin{align}\label{expectation}
	\langle \Omega | \phi (\tau) \ldots \phi (\tau) |\Omega \rangle \equiv \langle 0 | \bar F(\tau, \tau_0) \phi (\tau, \mathbf x_1) \ldots \phi (\tau, \mathbf x_N) F(\tau,\tau_0) | 0 \rangle~,
\end{align}
where $\tau$ is the time of the slice we want to evaluate the correlation on, and $\tau_0$ is the initial time. $F(\tau, \tau_0)$ is the usual evolution operator in the interaction picture
\begin{align}
	F(\tau,\tau_0) = {\rm T} \exp \bigg(-i \int_{\tau_0}^{\tau} d \tau_1 H_I (\tau_1) \bigg)~,
\end{align}
and $\bar F(\tau, \tau_0)$ is its Hermitian conjugate, where $\rm T$ denotes time ordering operator and $H_I(\tau)$ is the Hamiltonian in the interaction picture. 

In the Schwinger-Keldysh formalism, the expectation value \eqref{expectation} can be evaluated by inserting an identity operator $1 = \sum_\alpha | O_\alpha\rangle \langle O_\alpha|$ on the time slice at $\tau$ into the expectation value
\begin{align}
	\langle \Omega | \phi (\tau, \mathbf x_1) \ldots \phi (\tau, \mathbf x_N) | \Omega \rangle = \sum_{\alpha} \langle \Omega |  O_{\alpha} \rangle \langle O_{\alpha} |   \phi (\tau, \mathbf x_1) \ldots \phi (\tau, \mathbf x_N)  | \Omega \rangle~.
\end{align}
Here we have inserted the identity operator on the left of all the field operators, but it turns out that the result will be the same even if we had inserted it into somewhere else. Now we have some in-out inner products in this expression, we can write them as path integrals. To do this we insert the complete eigenbasis $1=\sum_{\phi(\tau_i)}|\phi(\tau_i) \rangle \langle\phi(\tau_i)|$ of the field operator and the complete eigenbasis $1=\sum_{\pi(\tau_i)}|\pi(\tau_i) \rangle \langle\pi(\tau_i)|$ of the conjugate momenta at each time slice $\tau_i$ for both of the time-ordered and the anti time-ordered factors. After integrating out the conjugate momenta, one arrives at the following formula
\begin{align}\nonumber
	\langle \phi (\tau, \mathbf x_1) \ldots \phi (\tau, \mathbf x_N) \rangle & = \int \mathcal D \phi_+   \mathcal D \phi_- \phi_+ (\tau, \mathbf x_1) \ldots \phi_+ (\tau, \mathbf x_N) \exp \bigg[ i \int_{\tau_0}^{\tau} d\tau' d^3{\bf x} (\mathcal L[\phi_+] - \mathcal L[\phi_-] ) \bigg] \delta\Big(\phi_+(\tau,\mathbf x) - \phi_- (\tau,\mathbf x) \Big)~,
\end{align} 
where $+$ and $-$ indicates the time-ordered and anti time-ordered factors, respectively. 

We define the generating functional as
\begin{align}
	Z[J_+, J_-] = \int \mathcal D \phi_+ \mathcal D \phi_- \exp \left[i \int_{\tau_0}^{\tau} d\tau' d^3 \mathbf x (\mathcal L[\phi_+] - \mathcal L[\phi_-] + J_+ \phi_+ - J_- \phi_- ) \right]  ~,
\end{align}
where $J_+(\tau,\mathbf x)$ and $J_-(\tau,\mathbf x)$ are sources for the scalar fields $\phi_+(\tau, \mathbf x)$ and $\phi_-(\tau, \mathbf x)$, respectively. The correlation function can be evaluated as
\begin{align}
	\langle \phi_{a_1} (\tau,\mathbf x_1) \ldots \phi_{a_N} (\tau,\mathbf x_N) \rangle = \frac{\delta}{i a_1 \delta J_{a_1} (\tau, \mathbf x_1) } \ldots \frac{\delta}{i a_N \delta J_{a_N} (\tau, \mathbf x_N) } Z[J_+, J_-] \bigg|_{J_{\pm} = 0}~,
\end{align}
where $a_i = \pm$. To do perturbative calculations, we split the Lagrangian density into free part $\mathcal L_0$ and interacting part $\mathcal L_{\rm int}$ as usual
\begin{align}
	\mathcal L[\phi] = \mathcal L_0 [\phi] +\mathcal L_{\rm int} [\phi]~,
\end{align} 
so that the generating functional can be written as
\begin{align}
	Z[J_+, J_-] &= \exp \bigg[i \int_{\tau_0}^{\tau_f} d\tau  d^3 \mathbf x \bigg(\mathcal L_{\rm int} \bigg[\frac{\delta}{i \delta J_+} \bigg] -\mathcal L_{\rm int} \bigg[ - \frac{\delta}{i \delta J_-}  \bigg] \bigg) \bigg] Z_0 [J_+, J_-]~, \\
	Z_0[J_+, J_-] & \equiv \int \mathcal D \phi_+ \mathcal D \phi_- \exp \bigg[i \int_{\tau_0}^{\tau_f}  d \tau d^3 \mathbf x \bigg(\mathcal L_0[\phi_+] -\mathcal L_0[\phi_-] + J_+ \phi_+ - J_-\phi_- \bigg) \bigg]~.
\end{align}
Then we can compute the correlation functions order by order. There are four types of propagators and they are defined as
\begin{align}
	- i \Delta_{ab} (\tau_1, \mathbf x, \tau_2, \mathbf y) = \frac{\delta }{i a \delta J_a (\tau_1,\mathbf x)}  \frac{\delta}{i b \delta J_b (\tau_2, \mathbf y)}  Z_0 [J_+, J_-] \bigg|_{J_{\pm} = 0}~.
\end{align}
We refer this kind of propagators as the \textit{bulk propagators}, or \textit{internal legs}. Usually it is more convenient to work with the Fourier transformation of it
\begin{align}
	G_{ab} (k; \tau_1, \tau_2) = - i \int d^3\mathbf x e^{-i \mathbf k\cdot (\mathbf x - \mathbf y)} \Delta_{ab} (\tau_1,\mathbf x; \tau_2, \mathbf y)~.
\end{align}
However, for our purpose, we will need a small modification: we need to make the momenta at the ``two sides" of the propagator different. For example, for a bulk propagator with momentum ${\bf l}_{ij}$ contains terms like $u(l_{ij},\tau_i)u^*(l_{ij},\tau_j)$, we will deform the momentum so that we have terms like $u(l_i,\tau_i)u^*(l_j,\tau_j)$, and set $l_i = l_j = l_{ij}$ at the end. To be more explicit, the four types of bulk propagators, written in terms of the mode functions of the fields, are
\begin{equation}
\begin{aligned}
	& G_{++} (l_1,\tau_1;l_2,\tau_2) = G_>(l_1,\tau_1;l_2,\tau_2) \theta(\tau_1- \tau_2) + G_< (l_1,\tau_1;l_2,\tau_2) \theta(\tau_2- \tau_1) \\
	& G_{+-} (l_1,\tau_1;l_2,\tau_2) = G_<(l_1,\tau_1;l_2,\tau_2) \\
	& G_{-+} (l_1,\tau_1;l_2,\tau_2) = G_>(l_1,\tau_1;l_2,\tau_2) \\
	& G_{--} (l_1,\tau_1;l_2,\tau_2) = G_<(l_1,\tau_1;l_2,\tau_2) \theta (\tau_1 - \tau_2) + G_>(l_1,\tau_1;l_2,\tau_2) \theta(\tau_2- \tau_1)
\end{aligned}
\end{equation}
where 
\begin{equation}
\begin{aligned}
	G_> (l_1,\tau_1;l_2,\tau_2) \equiv u_{l_1} (\tau_1) u^*_{l_2}(\tau_2) ~,\\
	G_< (l_1,\tau_1;l_2,\tau_2) \equiv u^*_{l_1} (\tau_1) u_{l_2} (\tau_2) ~,
\end{aligned}
\end{equation}
and $u_{l_i}$ are the mode functions of the fields. We will use these propagators with deformed momenta in this work, and impose the condition $l_1 = l_2 = l_{12}$ at the end. 

If $\tau_2$ is the time on the final time slice, $\tau_2 = \tau_f > \tau_1$, the propagators are called the \textit{bulk-to-boundary propagators}, or simply \textit{external legs}. For this kind of propagators, we will not need the trick of ``deforming the momenta" and we will have
\begin{equation}
\begin{aligned}
	 G_{+} (k,\tau_1) &\equiv G_{++} (k;\tau_1,\tau_f) =  G_{+-} (k;\tau_1,\tau_f) = G_< (k , \tau_1,\tau_f) \\
	 G_{-} (k,\tau_1) &\equiv G_{-+} (k;\tau_1,\tau_f) = G_{--} (k;\tau_1,\tau_f) = G_>(k , \tau_1,\tau_f) 
\end{aligned}
\end{equation}

If we use the perturbation theory to calculate the correlation functions, it is convenient to use diagrams to represent each terms in the perturbative expansion as in the usual quantum field theories. The rules for the diagrams are very similar to the usual Feynman rules in scalar field theories. We will emphasize the differences in this review.

For the external points, we will denote them using squares. Then for each vertex, we will differentiate them into two types: the {\it plus-type} and the {\it minus-type}, and we will denote them by black dots and white dots, respectively. Then the external leg will be represented as a line connecting two vertices, and the internal legs will be represented as a line connecting a vertex and an external point. Once we drawn the diagram, we can get an expression for the corresponding term in the perturbative expansion with the following rules:

For propagators,
\begin{equation}
\begin{aligned}
	\includegraphics[width=2cm]{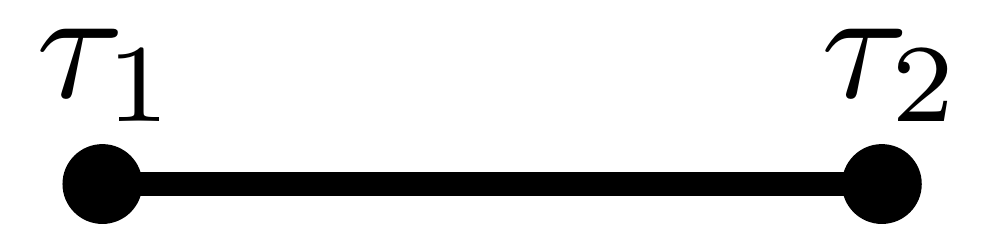} =\,\, & G_{++} (l_1,\tau_1;l_2,\tau_2) ~, \\
	\includegraphics[width=2cm]{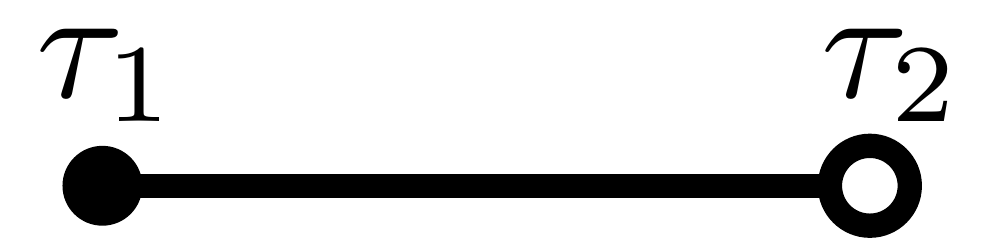} =\,\, & G_{+-} (l_1,\tau_1;l_2,\tau_2) ~,\\
	\includegraphics[width=2cm]{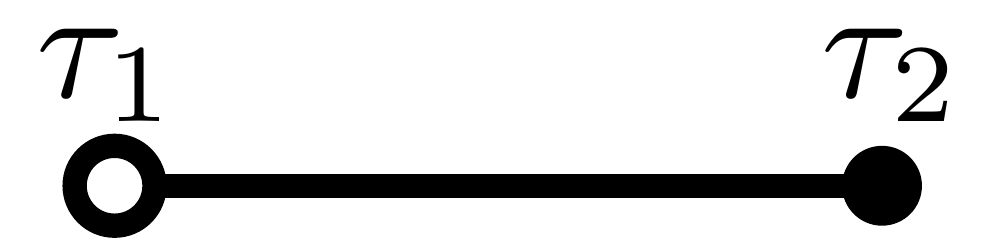} =\,\, & G_{-+} (l_1,\tau_1;l_2,\tau_2) ~,\\
	\includegraphics[width=2cm]{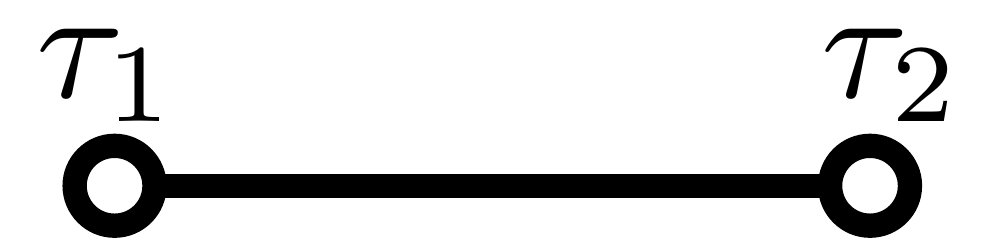} =\,\, & G_{--} (l_1,\tau_1;l_2,\tau_2) ~,\\
	\includegraphics[width=2cm]{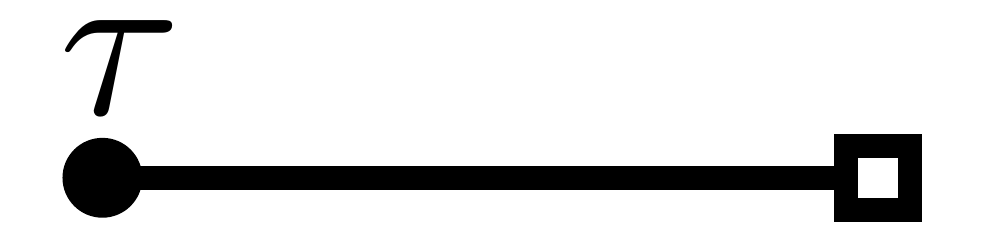} =\,\,& G_+(k;\tau)  ~,\\
	\includegraphics[width=2cm]{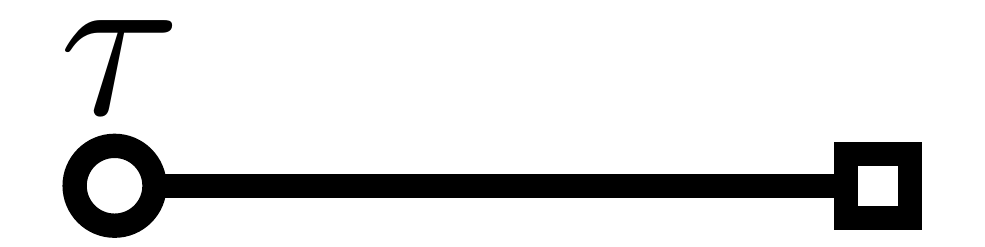} =\,\, & G_-(k;\tau) ~.
\end{aligned}
\end{equation}

For vertices of non-derivative coupling, {\it e.g.}, $\displaystyle {\cal L}_{\rm int} \supset -\frac{\lambda}{3!} a^4(\tau) \phi^3$, we have

\begin{equation}
	\includegraphics[width=3.5cm,valign=c]{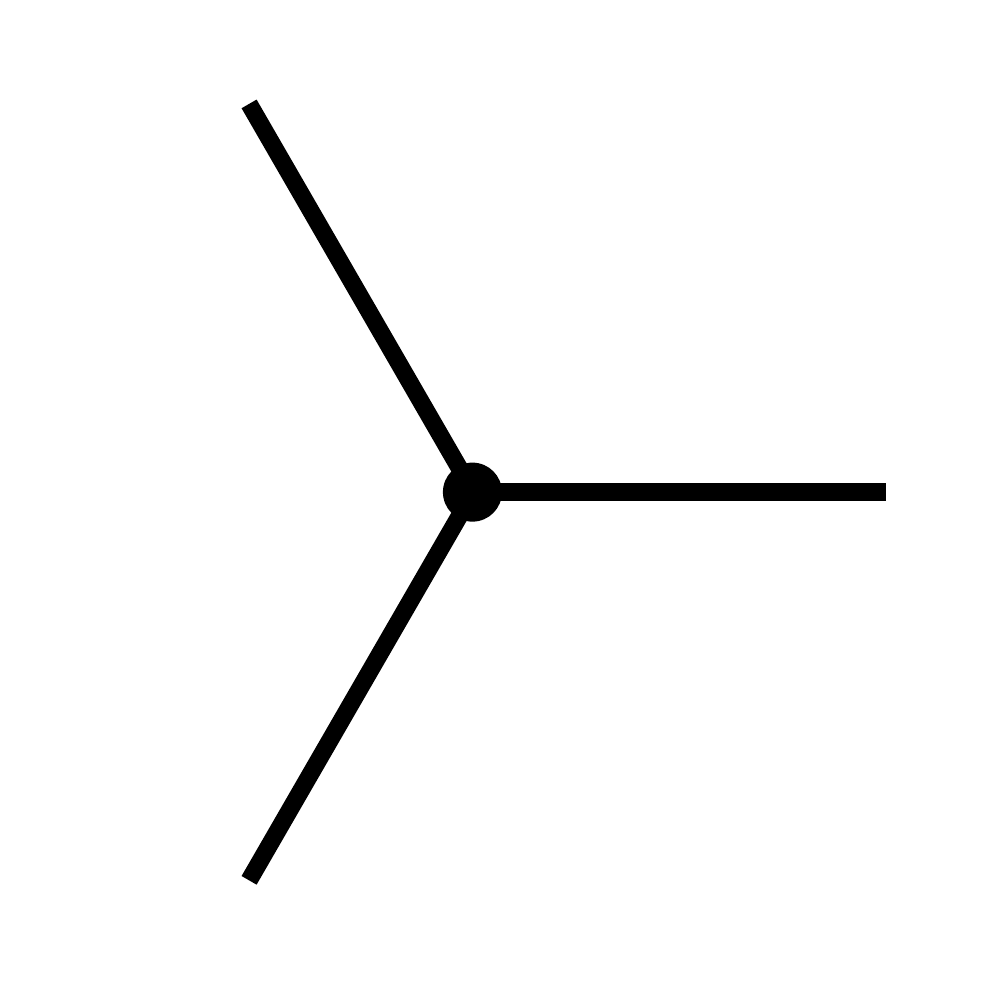} = - i \lambda \int_{\tau_0}^{\tau_f} d\tau ~ a^4(\tau) \cdots ~, \includegraphics[width=3.5cm,valign=c]{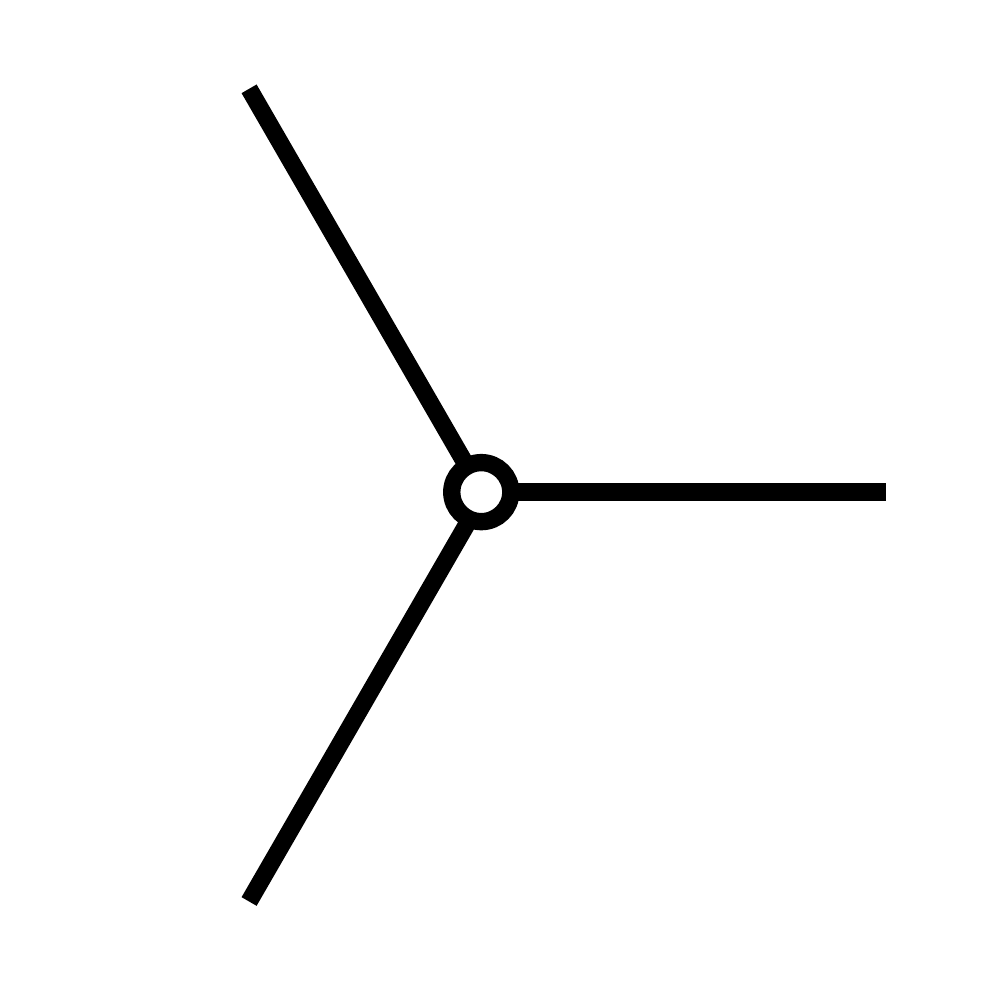} = + i \lambda \int_{\tau_0}^{\tau_f} d\tau ~ a^4(\tau) \cdots ~.
\end{equation}

For vertices with derivative coupling, we act the derivatives on the legs attached to the vertices: for spatial derivatives, we will get some factors of momentum, and for time derivatives we take the time derivatives of the propagators, {\it e.g.}, for $\displaystyle -\frac{\lambda}{2} a(\tau) \phi' \partial_i \phi \partial^i \phi$
\begin{equation}
	\includegraphics[width=3.5cm,valign=c]{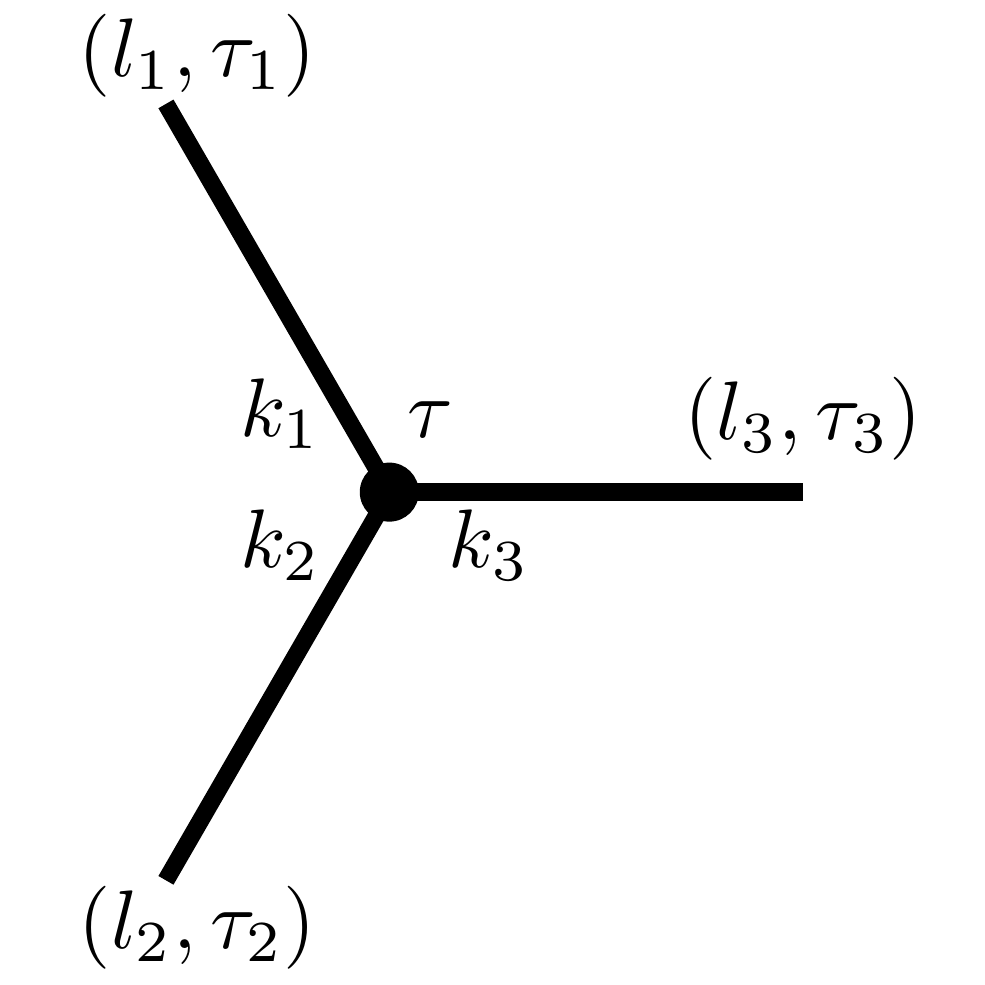} = + i \lambda ({\bf k}_1 \cdot {\bf k}_2) \int_{\tau_0}^{\tau_f} d\tau ~ a(\tau) G_{+a_1}(k_1,\tau ; l_1,\tau_1)G_{+a_2}(k_2,\tau ; l_2,\tau_2) \frac{\partial}{\partial \tau_3} G_{+a_3}(k_3,\tau ; l_3,\tau'_3) \cdots +\text{permutations}~,
\end{equation}
and for the diagram with a white dot instead of a black one, there is a minus sign in the expression. The sum of all diagrams gives us the expectation value with the overall momentum conserving delta function stripped off, {\it i.e.} the quantity $\langle \phi_{{\bf k}_1}\cdots\phi_{{\bf k}_n}\rangle'$, such that $\langle \phi_{{\bf k}_1}\cdots\phi_{{\bf k}_n}\rangle = (2\pi)^3 \delta^3 (\sum_i {\bf k}_i) \langle \phi_{{\bf k}_1}\cdots\phi_{{\bf k}_n}\rangle'$.

\section{Extracting operators from vertices}\label{operator}
In this section, we consider general diagrams for the curvature perturbation $\zeta$ (whose mode function takes the form of a massless scalar field in dS space), with cubic interactions $\displaystyle \frac{\kappa}{3!} a\zeta'^3$ and $\displaystyle \frac{\mu}{2} a\zeta'(\partial\zeta)^2$. Our goal is to construct an operator ${\cal O}_{\cal D}$ such that for a diagram ${\cal D}$ of $\zeta$, it equals to that operator acting on the same diagram, but with a massless scalar $\phi$ on the Minkowski space with $\displaystyle \frac{\lambda}{3!}\phi^3$ coupling,
\begin{equation}
\langle \zeta_{{\bf k}_1} \cdots \zeta_{{\bf k}_n} \rangle'_{\cal D} = {\cal O}_{\cal D} \langle \phi_{{\bf k}_1} \cdots \phi_{{\bf k}_n} \rangle'_{\cal D} ~,
\end{equation}
and both $n$-point functions are evaluated at $\tau = 0$. In this section we will introduce how can we get such an operator. And in appendix \ref{op_construction}, we will show that for each vertex there is a differential operator associated to it, independent of the ways of contractions, and whether the legs attached to the vertices being internal or external. Also, since we will take the momentum of the internal legs to be different at the two ends, and the operators associated to each vertex will only be depend on the momenta of the legs attached to it, the operators from each vertex will be independent and so they can commute with the propagators that are not attached to the relevant vertex. Therefore, we can pull these operators to the front of the whole diagram and get a complete operator that transform the diagram in flat space into the diagram in dS space. The result will be of the form
\begin{equation}
\langle \zeta_{{\bf k}_1} \cdots \zeta_{{\bf k}_n} \rangle'_{\cal D} = \left(\prod_i {{\cal O}_i} \right) \langle \phi_{{\bf k}_1} \cdots \phi_{{\bf k}_n} \rangle'_{\cal D} ~,
\end{equation}
where the index $i$ counts the vertices. 
Throughout this article, we will use $G$ and $u_k$ to denote propagators and mode functions for massless scalar field in the Minkowski space, and in de Sitter space we use $F$ and $v_k$. The mode functions are the solutions of the equation of motion (the Klein-Gordon equation) in the Fourier space. In the Minkowski space, it is simply
\begin{equation}
u_k'' (\tau) + k^2 u_k (\tau) = 0 ~,
\end{equation}
and its solution is 
\begin{equation}
u_k (\tau) = \frac{1}{\sqrt{2k}} e^{-i k \tau} ~.
\end{equation}
The coefficient $1 / \sqrt{2k}$ is fixed by the usual normalization condition:
\begin{equation}
u_k u_k^{*\prime} - u_k^* u_k' = i ~.
\end{equation}
In de Sitter space, the equation of motion is
\begin{equation}
(a v_k)'' + \left(k^2 - \frac{a''}{a} \right) (a v_k) = 0 ~,
\end{equation}
The solution is 
\begin{equation}
v_k (\tau) = \frac{H}{\sqrt{2k^3}} (1+i k \tau) e^{-i k \tau} ~.
\end{equation}
The coefficient $H / \sqrt{2k^3}$ is fixed by the condition:
\begin{equation}
a^2 (v_k v_k^{*\prime} - v_k^* v_k') = i ~.
\end{equation}
In both dS and Minkowski space, we focus on the Bunch-Davis vacuum.
For our formalism, we first strip those coefficients, {\it i.e}, we will write $u_k = e^{-ik\tau}$ and $v_k=(1+ik\tau) e^{-ik\tau}$, which can be easily restored. Also we set the Hubble parameter to be $H=1$ for convenience.

\subsection{The Method for Extracting the Operators}
Consider a general diagram ${\cal D}$ with the $i$-th vertex being a $a\zeta'^3$. The diagram will translate to the expression
\begin{equation}
\begin{aligned}
\langle\zeta_{{\bf k}_1}\zeta_{{\bf k}_2}\cdots\rangle'_{\cal D} = \sum_{\cal A}\int_{-\infty}^0 d\tau_i \int dT~\cdots \left(-\frac{1}{\tau_i}\right) \frac{\partial}{\partial \tau_i} \left[ {\cal V}_j F_{a_i a_j}(l_i,\tau_i; l_j,\tau_j) \right] \frac{\partial}{\partial \tau_i} \left[ {\cal V}_k F_{a_i a_k}(l'_i,\tau_i; l_k,\tau_k) \right] \frac{\partial}{\partial \tau_i} \left[ {\cal V}_l F_{a_i a_l}(l''_i,\tau_i; l_l,\tau_l) \right]\cdots~,
\end{aligned}
\end{equation}
where $\displaystyle \sum_{\cal A}$ means $\sum_{a_i=\pm} \left( \prod_i a_i \right)$, $\int dT$ denotes collectively all other time integrals, and ${\cal V}_j$, ${\cal V}_k$, ${\cal V}_l$ are the factors due to the possible space or time derivative from the vertex at time $\tau_j$, $\tau_k$, $\tau_l$, respectively. The method we use is rather simple, but sometimes powerful: we replace every $\tau_i$ in the integral by $\displaystyle \pm i K \frac{\partial}{\partial K}$, where $K$ is any one of the momenta attached to the vertex, {\it i.e.} $l_i$, $l'_i$ or $l''_i$. Since now we have a derivative with respect to the momenta, which is independent on time, we can pull this derivative out of the integral. Then we will see that the integral left is just the expression for exactly the same diagram we are considering, but is in the flat space with direct interaction. 

Now we illustrate how it works. From the above integral we can see that the quantity relevant to the $a\zeta'^3$ vertex will be 
\begin{equation}
\begin{aligned}
\left(-\frac{1}{\tau_i}\right) \left( \frac{\partial}{\partial \tau_i} v_{l_i} (\tau_i) \right) \left( \frac{\partial}{\partial \tau_i} v_{l'_i} (\tau_i) \right) \left( \frac{\partial}{\partial \tau_i} v_{l''_i} (\tau_i) \right) &= \left(-\frac{1}{\tau_i}\right) \left( \frac{\partial}{\partial \tau_i} (1+i l_i \tau_i) e^{-i l_i \tau_i} \right) \left( \frac{\partial}{\partial \tau_i} (1+i l'_i \tau_i) e^{-i l'_i \tau_i} \right) \left( \frac{\partial}{\partial \tau_i} (1+i l''_i \tau_i) e^{-i l''_i \tau_i} \right)\\
&= \left(-\frac{1}{\tau_i}\right) \left( l_i^2 \tau_i e^{-i l_i \tau_i} \right) \left( {l'_i}^2 \tau_i e^{-i l'_i \tau_i} \right) \left( {l''_i}^2 \tau_i e^{-i l''_i \tau_i} \right)\\
&= - l_i^2 {l'_i}^2 {l''_i}^2 \tau_i^2 e^{-i (l_i + l'_i + l''_i) \tau_i}
\end{aligned}
\end{equation}
and also its complex conjugate. Now we can replace the $\tau_i^2$ with $\displaystyle -K\frac{\partial}{\partial K}$, then we can pull the whole thing except the exponential out of the time integral. The only thing left inside the time integral is now $e^{-i (l_i + l'_i + l''_i) \tau_i} = u_{l_i} (\tau_i) u_{l'_i} (\tau_i) u_{l''_i} (\tau_i)$, which is associated to the $\phi^3$ vertex in flat space. So the operator associated to the $a\zeta'^3$ vertex is
\begin{equation}
\boxed {a\zeta'^3 :~ k_1^2 k_2^2 k_3^2 \frac{\partial^2}{\partial K^2}~,}
\end{equation}

This procedure can be done for any other kinds of interactions. For example, for the $a\zeta'(\partial\zeta)^2$ vertex, we have
\begin{equation}
\begin{aligned}
\left(-\frac{1}{\tau_i}\right) \left( {\bf l}_i \cdot {\bf l}'_i \right) v_{l_i} (\tau_i) v_{l'_i} (\tau_i) \left( \frac{\partial}{\partial \tau_i} v_{l''_i} (\tau_i) \right) &= \left(-\frac{1}{\tau_i}\right) \left( {\bf l}_i \cdot {\bf l}'_i \right) [(1+i l_i \tau_i) u_{l_i} (\tau_i)] [(1+i l'_i \tau_i) u_{l'_i} (\tau_i) ][{l''_i}^2 \tau_i u_{l''_i} (\tau_i)] ~.
\end{aligned}
\end{equation}
Then we replace $(1+i l_i \tau_i)$ with $\displaystyle \left(1- l_1 \frac{\partial}{\partial l_1}\right)$ and $(1+i l'_i \tau_i)$ with $\displaystyle \left(1- l'_1 \frac{\partial}{\partial l'_1}\right)$. So the operator associated to the $a\zeta'(\partial\zeta)^2$ vertex is
\begin{equation}
\boxed{a\zeta'(\partial\zeta)^2 :~ ({\bf k}_1\cdot{\bf k}_2) k_3^2 \left(1-k_1\frac{\partial}{\partial k_1}\right) \left(1-k_2\frac{\partial}{\partial k_2}\right) ~,}
\end{equation}

\section{A Recursion Relation for Tree-Level Correlation Functions in the Minkowski Space}\label{Rec}
\subsection{A Simple Example: Exchange Diagram}\label{ExchangeRec}
We start with a simple example: consider the s-channel exchange diagram for a 4-point function with a massless $\phi^3$ theory, evaluated at $\tau=0$,

\begin{figure}[htbp]
   \centering
   \includegraphics[scale=.5]{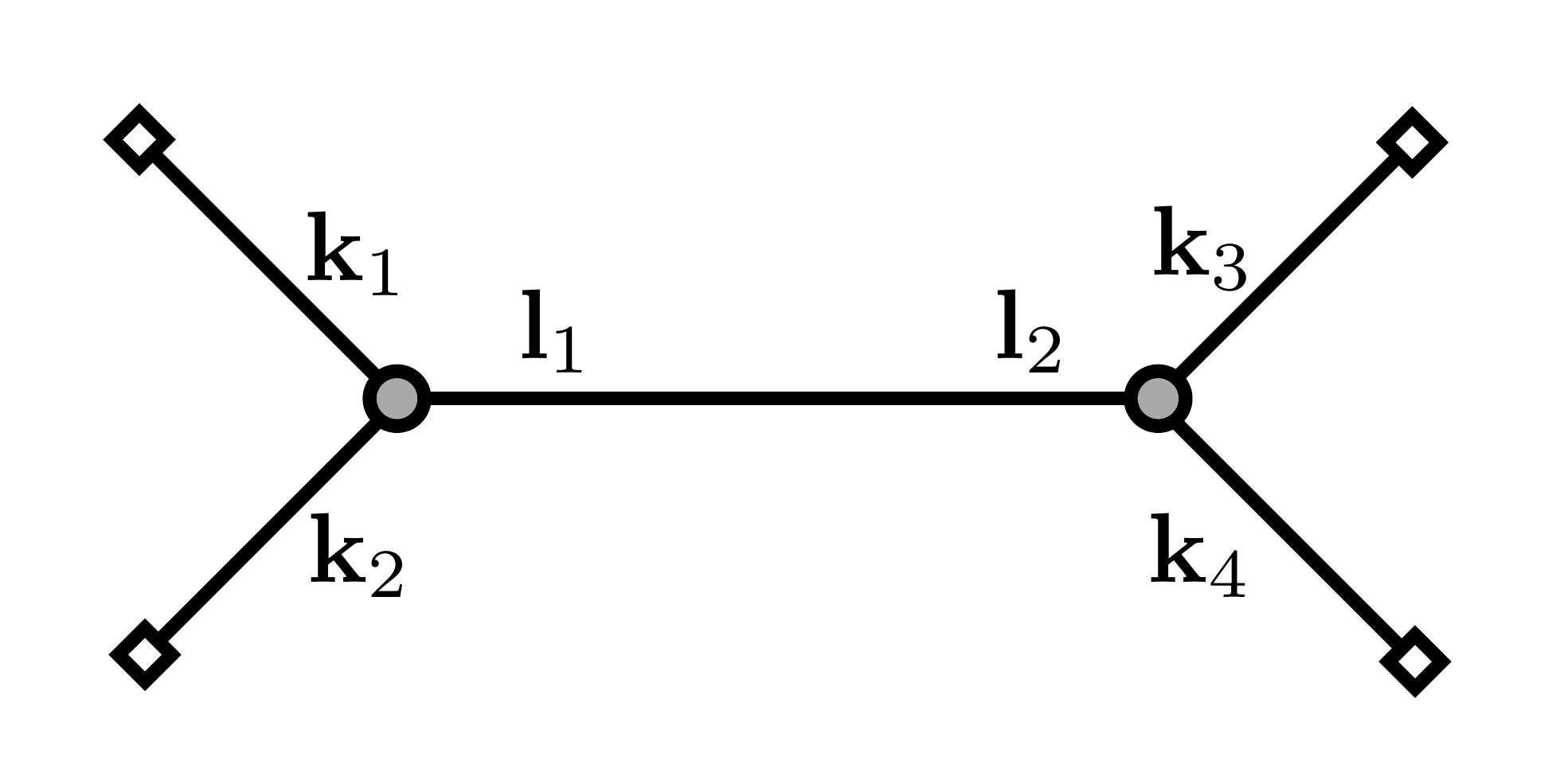} 
   \caption{The tree level diagram for 4-point function in $\phi^3$ theory.}
   \label{exchange}
\end{figure}

The grey dots are $+$ or $-$ type vertices. The contribution from this diagram is
\begin{equation}\label{2Site}
\begin{aligned}
\langle\phi_{{\bf k}_1}\phi_{{\bf k}_2}\phi_{{\bf k}_3}\phi_{{\bf k}_4}\rangle'_s=-\lambda^2 \sum_{\mathcal A} \int_{-\infty}^0d\tau_1\int_{-\infty}^0d\tau_2 \big[&G_{a_1}(k_1;\tau_1)G_{a_1}(k_2;\tau_1)G_{a_1a_2}(l_1,\tau_1 ; l_2,\tau_2)G_{a_2}(k_3;\tau_2)G_{a_2}(k_4;\tau_2) \big]
\end{aligned}
\end{equation} 
the subscript $s$ indicates this is the $s$-channel contribution only. The propagators are
\begin{equation}\label{prop_flat}
\begin{aligned}
G_{++}(l_1,\tau_1 ; l_2,\tau_2) = G_{--}^*(l_1,\tau_1 ; l_2,\tau_2)&=e^{-i (l_1\tau_1 - l_2\tau_2)}\theta(\tau_1-\tau_2)+e^{i ( l_1 \tau_1 - l_2 \tau_2 )}\theta(\tau_2-\tau_1)\\
G_{-+}(l_1,\tau_1 ; l_2,\tau_2)=G_{+-}^*(l_1,\tau_1 ; l_2,\tau_2)&=e^{-i ( l_1\tau_1 - l_2\tau_2)}\\
G_+(k;\tau)=G_-^*(k;\tau)&=e^{ik\tau}
\end{aligned}
\end{equation} 
We can also write the bulk propagator $G_{++}$ as 
\begin{equation}
G_{++}(l_1,\tau_1 ; l_2, \tau_2)=-\frac{1}{\pi i}\int_{-\infty}^{\infty}\frac{e^{i\omega(l_1\tau_1-l_2\tau_2)}}{\omega^2-1+i\epsilon}\,d\omega
\end{equation} 
For $G_{--}$, we simply take the complex conjugate. This equality is in fact not exact: it is only true after we impose the condition $l_1 = l_2 = l_{12}$, and before imposing it we will get step functions $\theta(l_i \tau_i-l_j \tau_j)$ instead of $\theta(\tau_i-\tau_j)$. However as long as there is no derivatives with respect to the $l$'s, everything will be fine after we take $l_1 = l_2 = l_{12}$. 

Consider the first two terms in \eqref{2Site}. Performing the $\tau_1$ integral, we get
\begin{equation}\label{2Site_partial}
\begin{aligned}
&\quad \int_{-\infty}^0 d\tau_1 \left[ G_+(k_1;\tau_1) G_+(k_2;\tau_1) G_{++}(l_1,\tau_1 ; l_2,\tau_2) - G_-(k_1;\tau_1) G_-(k_2;\tau_1) G_{-+}(l_1,\tau_1 ; l_2,\tau_2)\right] G_+(k_3;\tau_2) G_+(k_4;\tau_2)\\
&=\left[ -\frac{1}{\pi i} \int_{-\infty}^{\infty} d\omega \int_{-\infty}^0d\tau_1 \, e^{i(k_1+k_2)\tau_1} \frac{e^{i\omega(l_1\tau_1-l_2\tau_2)}}{\omega^2-1+i\epsilon} - \int_{-\infty}^0d\tau_1\, e^{-i(k_1+k_2)\tau_1} e^{-i(l_1\tau_1 - l_2\tau_2)}\right] e^{i(k_3+k_4)\tau_2}\\
&=\left[ \frac{1}{\pi} \int_{-\infty}^{\infty} d\omega \, \frac{1}{\omega l_1 - (-(k_1+k_2)+i\epsilon)} \frac{e^{-i\omega l_2 \tau_2}}{\omega^2-1+i\epsilon}-\frac{i}{(k_1+k_2)+l_1} e^{il_2\tau_2}\right] e^{i(k_3+k_4)\tau_2}\\
&=i\left[2\left(\lim_{\omega\rightarrow -(k_1+k_2)/l_1} \frac{e^{-i\omega l_2 \tau_2}}{l_1 (\omega^2 - 1)}+\lim_{\omega\rightarrow -1}\frac{1}{\omega l_1 + (k_1+k_2)}\frac{e^{-i\omega l_2 \tau_2}}{\omega-1}\right)-\frac{e^{i l_2\tau_2}}{(k_1+k_2)+l_ 1}\right]e^{i(k_3+k_4)\tau_2}\\
&=i\left[2\left(\frac{l_1 e^{i(k_1+k_2)(l_2/l_1)\tau_2}}{(k_1+k_2)^2-l_1^2}-\frac{1}{(k_1+k_2)-l_1}\frac{e^{il_2\tau_2}}{2}\right)-\frac{e^{i l_2 \tau_2}}{(k_1+k_2)+l_1}\right]e^{i(k_3+k_4)\tau_2}\\
&=\frac{2i}{(k_1+k_2)^2-l_1^2}\left[l_1e^{i((k_1+k_2)(l_2/l_1)+k_3+k_4)\tau_2}-(k_1+k_2)e^{i (l_2+k_3+k_4)\tau_2}\right]
\end{aligned}
\end{equation}

which means that a 4-point function can be reduced to a sum of some 3-point functions (with appropriate weightings), if we consider $\displaystyle (k_1+k_2)\frac{l_2}{l_1}$ as the magnitude of the momentum of an external leg in an unphysical 3-point function (since the momentum is not conserved in this 3-point function). For this reason, we will call this $\tilde{k}_{12}$ from now on, while we leave the symbol $k_{12}$ for $k_1+k_2$. The third and forth terms in \eqref{2Site} simply give us the complex conjugate of \eqref{2Site_partial}. Finally we do the $\tau_2$ integral and the result is
\begin{equation}
\langle\phi_{{\bf k}_1}\phi_{{\bf k}_2}\phi_{{\bf k}_3}\phi_{{\bf k}_4}\rangle'_s = -\frac{4\lambda^2}{ k_{12}^2-l_1^2}\left( \frac{l_1}{\tilde{k}_{12}+k_3+k_4} - \frac{k_{12}}{l_2+k_3+k_4} \right) ~.
\end{equation}
It is easy to check that it is the same as the result by directly integrating \eqref{2Site}. Notice that the three-point function with external momenta ${\bf k}_1$, ${\bf k}_2$ and ${\bf k}_3$ is $\langle\phi_{{\bf k}_1}\phi_{{\bf k}_2}\phi_{{\bf k}_3}\rangle' = -2\lambda/(k_1+k_2+k_3)$, so the equation above can also be written as 
\begin{equation}
\langle\phi_{{\bf k}_1}\phi_{{\bf k}_2}\phi_{{\bf k}_3}\phi_{{\bf k}_4}\rangle'_s = \frac{2\lambda}{k_{12}^2-l_1^2}\left( l_1 \langle\phi_{\tilde{{\bf k}}_{12}}\phi_{{\bf k}_3}\phi_{{\bf k}_4}\rangle' - k_{12}\langle\phi_{{\bf l}_{2}}\phi_{{\bf k}_3}\phi_{{\bf k}_4}\rangle' \right)
\end{equation}
where $|\tilde{{\bf k}}_{12}|=\tilde{k}_{12}$.

\subsection{General Diagrams}\label{GenDiag}
The above recursion relation can be easily generalized to any tree-level diagrams. In fact, even if the diagram contains loops, we still have the same recursion for every vertices that have only one internal leg attached to it, and this process can be iterated until there are no such vertices left. To show this, we consider a general diagram ${\cal D}$ with $n$ vertices, and consider one of such vertices. 
\begin{figure}[htbp]
   \centering
   \includegraphics[scale=.5]{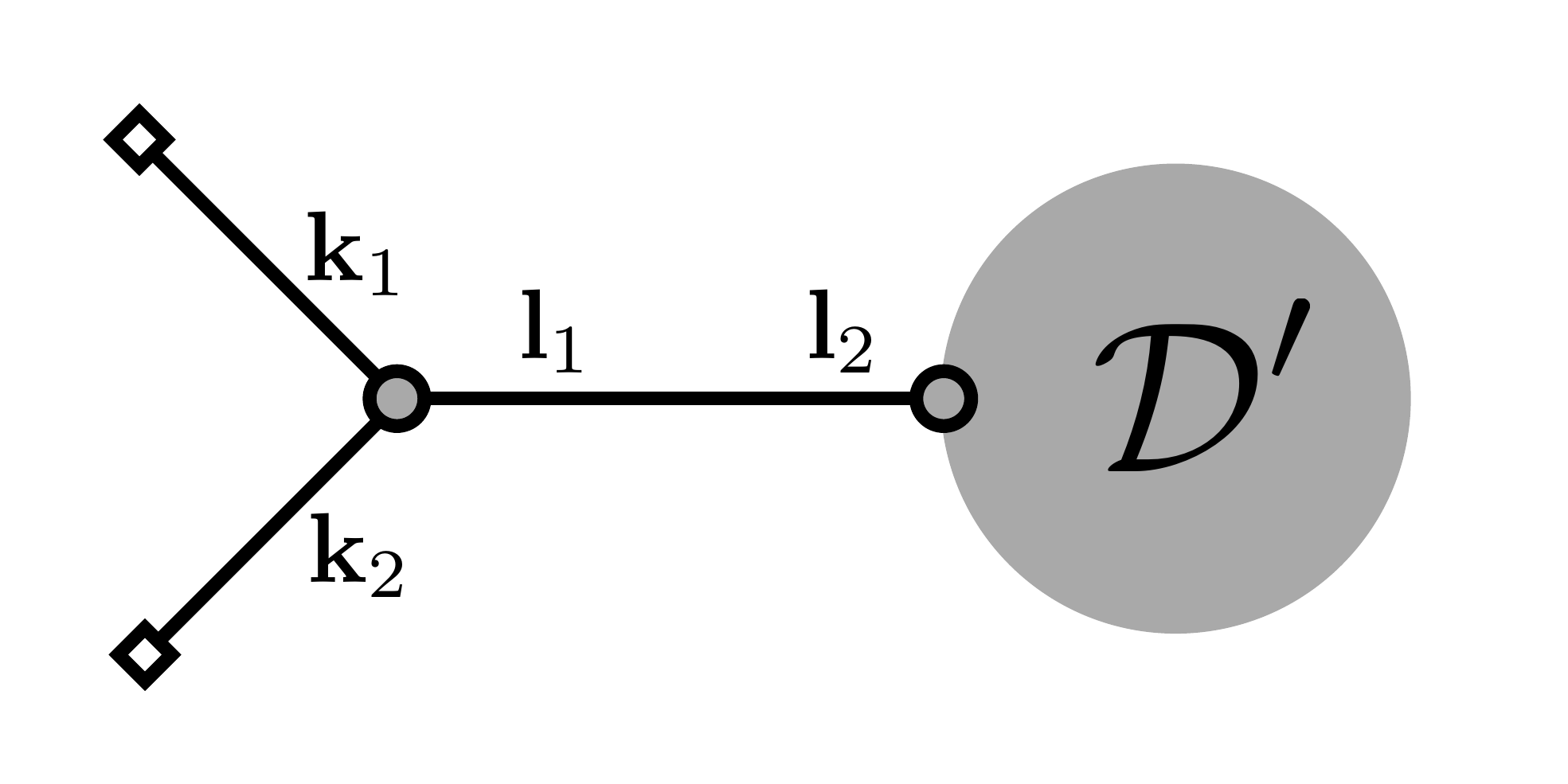} 
   \caption{A general diagram with a vertices that have only one internal leg attached to it. }
   \label{GeneralDiagram}
\end{figure}

This diagram represents
\begin{equation}
\begin{aligned}
\langle\phi_{{\bf k}_1}\phi_{{\bf k}_2}\cdots\rangle'_{\cal D} = -i \lambda \sum_{a_1=\pm} a_1 \bigg[ & \int_{-\infty}^0 d\tau_1 \int_{-\infty}^0 d\tau_2 \; G_{a_1}(k_1;\tau_1) G_{a_1}(k_2;\tau_1) G_{a_1+}(l_1,\tau_1 ; l_2,\tau_2){\cal D}'_+\\
- & \int_{-\infty}^0 d\tau_1 \int_{-\infty}^0 d\tau_2 \; G_{a_1}(k_1;\tau_1) G_{a_1}(k_2;\tau_1) G_{a_1-}(l_1,\tau_1 ; l_2,\tau_2){\cal D}'_- \bigg]
\end{aligned}
\end{equation}
where ${\cal D}'_\pm$ is the value of ${\cal D}'$ with the vertex at $\tau=\tau_2$ being $+$ and $-$, respectively. Schematically,
\begin{equation}
\begin{aligned}
{\cal D}'_\pm = (-i\lambda)^{n-1} \sum_{\cal A'} \int_{-\infty}^0 d\tau_3 \cdots \int_{-\infty}^0 d\tau_n \prod_{i \geq 2} G_{a_i} \prod_{j,k\geq2} G_{a_j a_k} ~,
\end{aligned}
\end{equation}
where $\displaystyle \prod_{i \geq 2} G_{a_i}$ means the product of the external legs and $\displaystyle \prod_{j,k \geq 2} G_{a_j a_k}$ means the product of the internal legs, and ${\cal A'} = {\cal A}-\{a_1\}$. 
Then following the derivation as in \eqref{2Site_partial}, we have
\begin{equation}
\begin{aligned}
\int_{-\infty}^0 d\tau_1 \int_{-\infty}^0 d\tau_2 \; G_{a_1}(k_1;\tau_1) G_{a_1}(k_2;\tau_1) G_{a_1\pm}(l_1,\tau_1 ; l_2,\tau_2){\cal D}'_\pm &= \frac{\pm 2i}{k_{12}^2-l_1^2} \int_{-\infty}^0 d\tau_2 \left(l_1 e^{\pm i\tilde{k}_{12}\tau_2}{\cal D}'_\pm - k_{12} e^{\pm i l_2 \tau_2}{\cal D}'_\pm \right) \\
&= \frac{\pm 2i}{k_{12}^2-l_1^2}\left(l_1 {\cal D}'_\pm(\tilde{k}_{12}) - k_{12} {\cal D}'_\pm(l_2)\right) ~,
\end{aligned}
\end{equation}
where ${\cal D}'_\pm(k) := \int d\tau_2 e^{\pm i k \tau_2}{\cal D}'_\pm$ is the value of the diagram with an external leg of momentum $k$ attached to ${\cal D}'_\pm$. Then we can write
\begin{equation}\label{GenRec}
\langle\phi_{{\bf k}_1}\phi_{{\bf k}_2}\cdots\rangle'_{\cal D} = \frac{2\lambda}{k_{12}^2-l_1^2} \left(l_1 \langle \phi_{\tilde{{\bf k}}_{12}} \cdots \rangle'_{\cal D'} - k_{12} \langle\phi_{{\bf l}_2} \cdots \rangle'_{\cal D'} \right) ~.
\end{equation}
For diagrams without loops, if we iterate this process over and over, the correlation function can be reduced to a linear combination of $2^n$ three-point functions.

\section{Recursion Relations in de Sitter Space}\label{RecDS}
With the operators constructed in Section \ref{operator}, we can translate the recursion relation obtained in Section \ref{Rec} by applying the operators on the both sides of \eqref{GenRec}. In this section, we demonstrate how to apply this method on the recursion relation of an exchange diagram in de Sitter space with $\displaystyle \frac{\kappa}{3!} a\zeta'^3$ and $\displaystyle \frac{\mu}{2} a\zeta'\partial\zeta\partial\zeta$ interactions. 

\subsection{$\displaystyle \frac{\kappa}{3!} a\zeta'^3$}
Recall that for a four-point function, we have
\begin{equation}
\langle\phi_{{\bf k}_1}\phi_{{\bf k}_2}\phi_{{\bf k}_3}\phi_{{\bf k}_4}\rangle'_s = \frac{2\lambda}{k_{12}^2-l_1^2}\left(l_1 \langle\phi_{{\bf k}_{12}}\phi_{{\bf k}_3}\phi_{{\bf k}_4}\rangle' - k_{12}\langle\phi_{{\bf l}_2}\phi_{{\bf k}_3}\phi_{{\bf k}_4}\rangle'\right) ~.
\end{equation}

For the diagram with both of the vertices being $a\zeta'^3$, we apply the operator $\displaystyle k_1^2 k_2^2 k_3^2 \frac{\partial^2}{\partial K^2}$ for both of the vertices. Since there are external legs attached to both vertices, we can choose the $K$ to be external momenta so that we can set $l_1=l_2=l_{12}$ first. The left hand side simply becomes the four-point function for $\zeta$ in dS. So we have
\begin{equation}
\begin{aligned}
\langle\zeta_{{\bf k}_1}\zeta_{{\bf k}_2}\zeta_{{\bf k}_3}\zeta_{{\bf k}_4}\rangle'_s = \left(k_1^2 k_2^2 l_{12}^2 \frac{\partial^2}{\partial k_1^2}\right) \left(k_3^2 k_4^2 l_{12}^2 \frac{\partial^2}{\partial k_3^2}\right) \left[ \frac{2\lambda}{k_{12}^2-l_{12}^2}\left(l_{12} \langle\phi_{{\bf k}_{12}}\phi_{{\bf k}_3}\phi_{{\bf k}_4}\rangle' - k_{12}\langle\phi_{{\bf l}_{12}}\phi_{{\bf k}_3}\phi_{{\bf k}_4}\rangle' \right) \right]  ~.
\end{aligned}
\end{equation}
For the right hand side, first, the operator $\displaystyle \left(k_3^2 k_4^2 l_{12}^2 \frac{\partial^2}{\partial k_3^2}\right)$ can be directly act on the three-point functions since the derivative is independent of $k_{12}$ and $l_{12}$. This gives
\begin{equation}
\begin{aligned}
k_3^2 k_4^2 l_{12}^2 \frac{\partial^2}{\partial k_3^2}\langle\phi_{{\bf k}_{12}}\phi_{{\bf k}_3}\phi_{{\bf k}_4}\rangle' &= \left(\frac{l_{12}^2}{k_{12}^2}\right) k_3^2 k_4^2 k_{12}^2 \frac{\partial^2}{\partial k_3^2}\langle\phi_{{\bf k}_{12}}\phi_{{\bf k}_3}\phi_{{\bf k}_4}\rangle' = \left(\frac{l_{12}^2}{k_{12}^2}\right) \frac{\lambda}{\kappa} \langle\zeta_{{\bf k}_{12}}\zeta_{{\bf k}_3}\zeta_{{\bf k}_4}\rangle' ~, \\
k_3^2 k_4^2 l_{12}^2 \frac{\partial^2}{\partial k_3^2}\langle\phi_{{\bf l}_{12}}\phi_{{\bf k}_3}\phi_{{\bf k}_4}\rangle' &= \frac{\lambda}{\kappa} \langle\zeta_{{\bf l}_{12}}\zeta_{{\bf k}_3}\zeta_{{\bf k}_4}\rangle' ~.
\end{aligned}
\end{equation}
Then we can express the final answer as derivatives of the three-point functions:
\begin{equation} 
\begin{aligned}
\langle\zeta_{{\bf k}_1}\zeta_{{\bf k}_2}\zeta_{{\bf k}_3}\zeta_{{\bf k}_4}\rangle'_s = \left(k_1^2 k_2^2 l_{12}^2 \frac{\partial^2}{\partial k_1^2}\right) \left[ \frac{2\lambda^2 / \kappa}{k_{12}^2-l_{12}^2}\left( \frac{l_{12}^3}{k_{12}^2} \langle\zeta_{{\bf k}_{12}}\zeta_{{\bf k}_3}\zeta_{{\bf k}_4}\rangle' - k_{12}\langle\zeta_{{\bf l}_{12}}\zeta_{{\bf k}_3}\zeta_{{\bf k}_4}\rangle' \right) \right]  ~.
\end{aligned}
\end{equation}

\subsection{$\displaystyle \frac{\mu}{2} a\zeta'\partial\zeta\partial\zeta$}
For the diagram with both of the vertices being $a\zeta'\partial\zeta\partial\zeta$, there are four different relevant diagrams (up to the permutations $k_1 \leftrightarrow k_2$ and $k_3 \leftrightarrow k_4$), which have different internal legs. The $\zeta'-\zeta'$ one, the $\zeta'-\partial\zeta$ one, the $\partial\zeta-\zeta'$ one, and the $\partial\zeta-\partial\zeta$ one. This time we apply the operator $\displaystyle \left[ ({\bf k}_1\cdot{\bf k}_2) k_3^2 \left(1-k_1\frac{\partial}{\partial k_1}\right) \left(1-k_2\frac{\partial}{\partial k_2}\right) \right]$ on both of the vertices:
\begin{align}
\zeta'-\zeta' &: ({\bf k}_1\cdot{\bf k}_2) ({\bf k}_3\cdot{\bf k}_4) l_1^2 l_2^2 \left(1-k_1\frac{\partial}{\partial k_1}\right) \left(1-k_2\frac{\partial}{\partial k_2}\right) \left(1-k_3\frac{\partial}{\partial k_3}\right) \left(1-k_4\frac{\partial}{\partial k_4}\right) ~,\\
\zeta'-\partial\zeta &: ({\bf k}_1\cdot{\bf k}_2) ({\bf k}_3\cdot{\bf l}_2) l_1^2 k_4^2 \left(1-k_1\frac{\partial}{\partial k_1}\right) \left(1-k_2\frac{\partial}{\partial k_2}\right) \left(1-k_3\frac{\partial}{\partial k_3}\right) \left(1-l_2\frac{\partial}{\partial l_2}\right) ~,\\
\partial\zeta-\zeta' &: ({\bf l}_1\cdot{\bf k}_2) ({\bf k}_3\cdot{\bf k}_4) k_1^2 l_2^2 \left(1-l_1\frac{\partial}{\partial l_1}\right) \left(1-k_2\frac{\partial}{\partial k_2}\right) \left(1-k_3\frac{\partial}{\partial k_3}\right) \left(1-k_4\frac{\partial}{\partial k_4}\right) ~,\\
\partial\zeta-\partial\zeta &: ({\bf l}_1\cdot{\bf k}_2) ({\bf k}_3\cdot{\bf l}_2) k_1^2 k_4^2 \left(1-l_1\frac{\partial}{\partial l_1}\right) \left(1-k_2\frac{\partial}{\partial k_2}\right) \left(1-k_3\frac{\partial}{\partial k_3}\right) \left(1-l_2\frac{\partial}{\partial l_2}\right) ~.
\end{align}
Again, the operator associated to the second vertex can be directly act on the three-point functions and give us dS three-point functions:
\begin{equation} 
\begin{aligned}
({\bf k}_3 \cdot {\bf k}_4) l_2^2 \left( 1- k_3\frac{\partial}{\partial k_3}\right) \left( 1- k_4\frac{\partial}{\partial k_4}\right) \langle\phi_{{\bf \tilde{k}}_{12}}\phi_{{\bf k}_3}\phi_{{\bf k}_4}\rangle' &= \frac{l_2^2}{\tilde{k}_{12}^2} ({\bf k}_3 \cdot {\bf k}_4) \tilde{k}_{12}^2 \left( 1- k_3\frac{\partial}{\partial k_3}\right) \left( 1- k_4\frac{\partial}{\partial k_4}\right) \langle\phi_{{\bf \tilde{k}}_{12}}\phi_{{\bf k}_3}\phi_{{\bf k}_4}\rangle' \\
&= \frac{l_1^2}{k_{12}^2} \frac{\lambda}{\mu} \langle \zeta^t_{{\bf \tilde{k}}_{12}}\zeta_{{\bf k}_3}\zeta_{{\bf k}_4}\rangle' ~, \\
({\bf k}_3 \cdot {\bf k}_4) l_2^2 \left( 1- k_3\frac{\partial}{\partial k_3}\right) \left( 1- k_4\frac{\partial}{\partial k_4}\right)\langle\phi_{{\bf l}_2}\phi_{{\bf k}_3}\phi_{{\bf k}_4}\rangle' &= \frac{\lambda}{\mu} \langle \zeta^t_{{\bf l}_2}\zeta_{{\bf k}_3}\zeta_{{\bf k}_4}\rangle' ~.
\end{aligned}
\end{equation}
where the superscript $t$ in $\zeta^t$ indicates that this field is contracted to the $\dot{\zeta}$ field from the vertex. That is, the $\langle\cdots\rangle$ on the right hand side are not the complete tree-level three point function, and in this notation it should be $\langle \zeta_{{\bf k}_1}\zeta_{{\bf k}_2}\zeta_{{\bf k}_3}\rangle'=\langle \zeta^t_{{\bf k}_1}\zeta_{{\bf k}_2}\zeta_{{\bf k}_3}\rangle'+\langle \zeta_{{\bf k}_1}\zeta^t_{{\bf k}_2}\zeta_{{\bf k}_3}\rangle'+\langle \zeta_{{\bf k}_1}\zeta_{{\bf k}_2}\zeta^t_{{\bf k}_3}\rangle'$. 

Then, for $\zeta'-\zeta'$ we get
\begin{equation} 
\begin{aligned}
\langle\zeta_{{\bf k}_1}\zeta_{{\bf k}_2}\zeta_{{\bf k}_3}\zeta_{{\bf k}_4}\rangle'_{\zeta'-\zeta'} = ({\bf k}_1\cdot{\bf k}_2) l_1^2 \left(1-k_1\frac{\partial}{\partial k_1}\right) \left(1-k_2\frac{\partial}{\partial k_2}\right) \left[ \frac{2\lambda^2 / \mu}{k_{12}^2-l_1^2}\left( \frac{l_1^3}{k_{12}^2} \langle\zeta^t_{{\bf k}_{12}}\zeta_{{\bf k}_3}\zeta_{{\bf k}_4}\rangle' - k_{12}\langle\zeta^t_{{\bf l}_2}\zeta_{{\bf k}_3}\zeta_{{\bf k}_4}\rangle' \right) \right] \Bigg|_{l_1=l_2=l_{12}} ~.
\end{aligned}
\end{equation}

For $\partial\zeta-\zeta'$ we have
\begin{equation} 
\begin{aligned}
\langle\zeta_{{\bf k}_1}\zeta_{{\bf k}_2}\zeta_{{\bf k}_3}\zeta_{{\bf k}_4}\rangle'_{\partial\zeta-\zeta'} = ({\bf l}_1\cdot{\bf k}_2) k_1^2 \left(1-l_1\frac{\partial}{\partial l_1}\right) \left(1-k_2\frac{\partial}{\partial k_2}\right) \left[ \frac{2\lambda^2 / \mu}{k_{12}^2-l_1^2}\left( \frac{l_1^3}{k_{12}^2} \langle\zeta^t_{{\bf k}_{12}}\zeta_{{\bf k}_3}\zeta_{{\bf k}_4}\rangle' - k_{12}\langle\zeta^t_{{\bf l}_2}\zeta_{{\bf k}_3}\zeta_{{\bf k}_4}\rangle' \right) \right] \Bigg|_{l_1=l_2=l_{12}} ~.
\end{aligned}
\end{equation}

For $\zeta'-\partial\zeta$, observe that 
\begin{equation}
l_2 \frac{\partial}{\partial l_2} \langle\phi_{{\tilde{\bf k}}_{12}}\phi_{{\bf k}_3}\phi_{{\bf k}_4}\rangle' = l_2\frac{\partial}{\partial l_2} f\left(k_{12} \frac{l_2}{l_1}\right) = \tilde{k}_{12}\frac{\partial}{\partial \tilde{k}_{12}} f(\tilde{k}_{12}) = \tilde{k}_{12}\frac{\partial}{\partial \tilde{k}_{12}} \langle\phi_{{\tilde{\bf k}}_{12}}\phi_{{\bf k}_3}\phi_{{\bf k}_4}\rangle' ~.
\end{equation}
Therefore we have
\begin{equation} 
\begin{aligned}
({\bf k}_3 \cdot {\bf l}_2) k_4^2 \left( 1- k_3\frac{\partial}{\partial k_3}\right) \left( 1- l_2 \frac{\partial}{\partial l_2}\right) \langle\phi_{{\bf \tilde{k}}_{12}}\phi_{{\bf k}_3}\phi_{{\bf k}_4}\rangle' &= \frac{{\bf k}_3 \cdot {\bf l}_2}{{\bf k}_3 \cdot {\bf \tilde{k}}_{12}} ({\bf k}_3 \cdot {\bf \tilde{k}}_{12}) k_4^2 \left( 1- k_3\frac{\partial}{\partial k_3}\right) \left( 1- \tilde{k}_{12} \frac{\partial}{\partial \tilde{k}_{12}}\right) \langle\phi_{{\bf \tilde{k}}_{12}}\phi_{{\bf k}_3}\phi_{{\bf k}_4}\rangle' \\
&= \left(\frac{l_1}{l_2}\right) \frac{{\bf k}_3 \cdot {\bf l}_2}{{\bf k}_3 \cdot {\bf k}_{12}} \frac{\lambda}{\mu} \langle \zeta_{{\bf \tilde{k}}_{12}} \zeta_{{\bf k}_3} \zeta^t_{{\bf k}_4}\rangle' ~, \\
({\bf k}_3 \cdot {\bf l}_2) k_4^2 \left( 1- k_3\frac{\partial}{\partial k_3}\right) \left( 1- l_2\frac{\partial}{\partial l_2}\right)\langle\phi_{{\bf l}_2}\phi_{{\bf k}_3}\phi_{{\bf k}_4}\rangle' &= \frac{\lambda}{\mu} \langle \zeta_{{\bf l}_2} \zeta_{{\bf k}_3} \zeta^t_{{\bf k}_4}\rangle' ~.
\end{aligned}
\end{equation}
Then we get, for $\zeta'-\partial\zeta$,
\begin{equation} 
\begin{aligned}
\langle\zeta_{{\bf k}_1}\zeta_{{\bf k}_2}\zeta_{{\bf k}_3}\zeta_{{\bf k}_4}\rangle'_{\zeta'-\partial\zeta} = ({\bf k}_1\cdot{\bf k}_2) l_1^2 \left(1-k_1\frac{\partial}{\partial k_1}\right) \left(1-k_2\frac{\partial}{\partial k_2}\right) \left[ \frac{2\lambda^2 / \mu}{k_{12}^2-l_1^2}\left( \frac{l_1^2}{l_2} \frac{{\bf k}_3 \cdot {\bf l}_2}{{\bf k}_3 \cdot {\bf k}_{12}}  \langle\zeta_{{\bf k}_{12}}\zeta_{{\bf k}_3}\zeta^t_{{\bf k}_4}\rangle' - k_{12}\langle\zeta_{{\bf l}_2}\zeta_{{\bf k}_3}\zeta^t_{{\bf k}_4}\rangle' \right) \right] \Bigg|_{l_1=l_2=l_{12}} ~.
\end{aligned}
\end{equation}

Finally, for $\partial\zeta-\partial\zeta$ we need to first notice that for the bulk propagator 
\begin{equation}\label{w-integral}
G_{++}(l_1,\tau_1 ; l_2, \tau_2) = -\frac{1}{\pi i}\int_{-\infty}^{\infty}\frac{e^{i\omega(l_1\tau_1-l_2\tau_2)}}{\omega^2-1+i\epsilon}\,d\omega ~,
\end{equation} 
not only the second time derivatives $\displaystyle \frac{\partial}{\partial \tau_1} \frac{\partial}{\partial \tau_2}$ will give a delta function, $\displaystyle \frac{\partial}{\partial l_1} \frac{\partial}{\partial l_2}$ can also give a delta function: 
\begin{equation}\label{subtract-delta}
\begin{aligned}
&\quad \left[\frac{\partial}{\partial l_1} \frac{\partial}{\partial l_2}G_{>}(l_1,\tau_1 ; l_2, \tau_2)\right]\theta(\tau_1-\tau_ 2) + \left[\frac{\partial}{\partial l_1} \frac{\partial}{\partial l_2}G_{<}(l_1,\tau_1 ; l_2, \tau_2)\right]\theta(\tau_2-\tau_1) \\
&= -\frac{1}{\pi i}\int_{-\infty}^{\infty}\frac{\omega^2 \tau_1 \tau_2 e^{i\omega(l_1\tau_1-l_2\tau_2)}}{\omega^2-1+i\epsilon}\,d\omega - \frac{2 \tau_1 \tau_2}{l_{12}} i \delta(\tau_1-\tau_2) ~.
\end{aligned}
\end{equation}
When we derive the recursion relation in Minkowski space, we used the $\omega$ integral in \eqref{w-integral} to replace the propagator, so this delta function in \eqref{subtract-delta} have to be subtracted manually from the recursion relation. Therefore, for the diagram with the $\partial\zeta-\partial\zeta$ internal propagator we have to subtract the induced four-point function with the interaction term ${\cal L} \supset 2 \lambda l_{12} \zeta'^2 \zeta^2$ and the result will be
\begin{equation} 
\begin{aligned}
\langle\zeta_{{\bf k}_1}\zeta_{{\bf k}_2}\zeta_{{\bf k}_3}\zeta_{{\bf k}_4}\rangle'_{\partial\zeta-\partial\zeta} &= ({\bf l}_1\cdot{\bf k}_2) k_1^2 \left(1-l_1\frac{\partial}{\partial l_1}\right) \left(1-k_2\frac{\partial}{\partial k_2}\right) \left[ \frac{2\lambda^2 / \mu}{k_{12}^2-l_1^2}\left( \frac{l_1^2}{l_2} \frac{{\bf k}_3 \cdot {\bf l}_2}{{\bf k}_3 \cdot {\bf k}_{12}}  \langle\zeta_{{\bf k}_{12}}\zeta_{{\bf k}_3}\zeta^t_{{\bf k}_4}\rangle' - k_{12}\langle\zeta_{{\bf l}_2}\zeta_{{\bf k}_3}\zeta^t_{{\bf k}_4}\rangle' \right) \right] \Bigg|_{l_1=l_2=l_{12}} \\
& \quad - ({\bf l}_{12}\cdot{\bf k}_2)({\bf l}_{12}\cdot{\bf k}_3) \langle \zeta^t_{{\bf k}_1} \zeta_{{\bf k}_2} \zeta_{{\bf k}_3} \zeta^t_{{\bf k}_4}\rangle'_{\rm Induced} ~.
\end{aligned}
\end{equation}
where $\langle \zeta^t_{{\bf k}_1} \zeta_{{\bf k}_2} \zeta_{{\bf k}_3} \zeta^t_{{\bf k}_4}\rangle'_{\rm Induced}$ is the four-point contact diagram with the interaction $2 \lambda l_{12} \zeta'^2 \zeta^2$ induced by the second momentum derivative. Note that derivatives like $\displaystyle \frac{\partial}{\partial l_1} \frac{\partial}{\partial \tau_1}$ can also produce a delta function, but when calculating correlation functions, they will not contribute to the final result if the late time limit is taken. 

Summing over the four different contractions, we obtain the full $s$-channel contribution of the four-point function due to the $\displaystyle \frac{\mu}{2} a\zeta'\partial\zeta\partial\zeta$ interaction,
\begin{equation} 
\begin{aligned}
\langle\zeta_{{\bf k}_1}\zeta_{{\bf k}_2}\zeta_{{\bf k}_3}\zeta_{{\bf k}_4}\rangle'_s &= \left[ ({\bf k}_1\cdot{\bf k}_2) l_1^2 \left(1-k_1\frac{\partial}{\partial k_1}\right) \left(1-k_2\frac{\partial}{\partial k_2}\right) + ({\bf k}_1 \leftrightarrow {\bf l}_1) \right] \left[ \frac{2\lambda^2 / \mu}{k_{12}^2-l_1^2}\left( \frac{l_1^3}{k_{12}^2} \langle\zeta^t_{{\bf k}_{12}}\zeta_{{\bf k}_3}\zeta_{{\bf k}_4}\rangle' - k_{12}\langle\zeta^t_{{\bf l}_2}\zeta_{{\bf k}_3}\zeta_{{\bf k}_4}\rangle' \right) \right] \Bigg|_{l_1=l_2=l_{12}}\\
& + \left[ ({\bf k}_1\cdot{\bf k}_2) l_1^2 \left(1-k_1\frac{\partial}{\partial k_1}\right) \left(1-k_2\frac{\partial}{\partial k_2}\right)  + ({\bf k}_1 \leftrightarrow {\bf l}_1) \right]  \left[ \frac{2\lambda^2 / \mu}{k_{12}^2-l_1^2}\left( \frac{l_1^2}{l_2} \frac{{\bf k}_3 \cdot {\bf l}_2}{{\bf k}_3 \cdot {\bf k}_{12}}  \langle\zeta_{{\bf k}_{12}}\zeta_{{\bf k}_3}\zeta^t_{{\bf k}_4}\rangle' - k_{12}\langle\zeta_{{\bf l}_2}\zeta_{{\bf k}_3}\zeta^t_{{\bf k}_4}\rangle' \right) \right] \Bigg|_{l_1=l_2=l_{12}} \\
& \quad - ({\bf l}_{12}\cdot{\bf k}_2)({\bf l}_{12}\cdot{\bf k}_3) \langle \zeta^t_{{\bf k}_1} \zeta_{{\bf k}_2} \zeta_{{\bf k}_3} \zeta^t_{{\bf k}_4}\rangle'_{\rm Induced} ~.
\end{aligned}
\end{equation}

One can also find the recursion relations for diagrams with both of these interactions, diagrams with more than two vertices, etc., and the procedures will be the same as above.

\section{Conclusion}
We developed the systematic way of constructing operators that relate the inflationary correlation function with Minkowski correlation function. Given a diagram and the form of the vertices, the corresponding operator can be directly written down. This formalism improved the generality, clarity and simplicity compared to our previous one developed in \cite{Chu:2018ovy}. 

We also discussed the recursion relation directly at the correlation level. In our previous work, we first find a recursion relation between higher-point and lower-point about the wave function of the universe and then using the recursion relation between the wave function to obtain the recursion relation of the correlation function in flat spacetime. Afterwards, we use the operator method to obtain the recursion relation in de Sitter space. In contrast, here we found that by using a similar technique developed in \cite{Arkani-Hamed:2017fdk}, we can actually find a recursion relation for the correlation function in flat space directly. By applying an appropriate operator, we can get the corresponding recursion relation in de Sitter space.

\section*{Acknowledgments}  
We thank Shingyan Li, Shiyun Lu and Xi Tong for useful discussions. This work is supported  by ECS Grant 26300316 and GRF Grant 16301917 from the Research Grants Council of Hong Kong. 

\appendix

\section{Constructions of the operators}\label{op_construction}
In this appendix we will show the details of the constructions of the operators corresponds to different vertices in different situation. 
\subsection{$\displaystyle \frac{\kappa}{3!} a\zeta'^3$ vertex}
Let's start with a $\displaystyle\frac{\kappa}{3!} a\zeta'^3$ vertex. We first note that the time derivatives of $F$ are
\begin{equation}
\begin{aligned}
\frac{\partial}{\partial \tau_i} F_{a_i}(k,\tau_i) &= k^2 \tau_i G_{a_i}(k,\tau_i) \\
\frac{\partial}{\partial \tau_i} F_{a_i a_j}(l_i,\tau_i ; l_j,\tau_j) &= l_i^2 \tau_i \tilde{F}_{a_i a_j}(l_i,\tau_i ; l_j,\tau_j) \\
&\quad + \delta_{a_i a_j}\left[ (1 + i l_i \tau_i) (1 - i l_j \tau_i) G_>(l_i,\tau_i ; l_j,\tau_j) - (1 - i l_i \tau_i) (1 + i l_j \tau_i) G_<(l_i,\tau_i ; l_j,\tau_j) \right] \delta(\tau_i-\tau_j)~.
\end{aligned}
\end{equation}
where $a_i=\pm$ and the $\sim$ symbol on $\tilde{F}_{a_i a_j}$ means the mode functions at time $\tau_i$ are replaced by the mode functions in flat spacetime, that is, 
\begin{equation}
\begin{aligned}
\tilde{F}_{>}(l_1,\tau_1 ; l_2,\tau_2) &:= u_{k_i}(\tau_i) v^*_{k_j}(\tau_j) = [e^{-i l_1 \tau_1}] [(1 - i l_2 \tau_2)e^{i l_2 \tau_2}] \\
\tilde{F}_{<}(l_1,\tau_1 ; l_2,\tau_2) &:= u^*_{k_i}(\tau_i) v_{k_j}(\tau_j) = [e^{i l_1 \tau_1}] [(1 + i l_2 \tau_2)e^{-i l_2 \tau_2}]~.
\end{aligned}
\end{equation}
Notice that, for $a_i = a_j$, the time derivative gives a term with $\delta$ function due to the Heaviside step function in the propagator. This term will vanish as we take the equal-momentum condition, $l_1 = l_2 = l_{12}$. However, if there is a second derivative on it, there will be one more extra term with a derivative of the $\delta$ function. This will not vanish and will induce higher order interactions \cite{Chen:2017ryl}.

\subsubsection{Three external legs}
\begin{figure}[htbp]
	\centering
	\includegraphics[scale=.5]{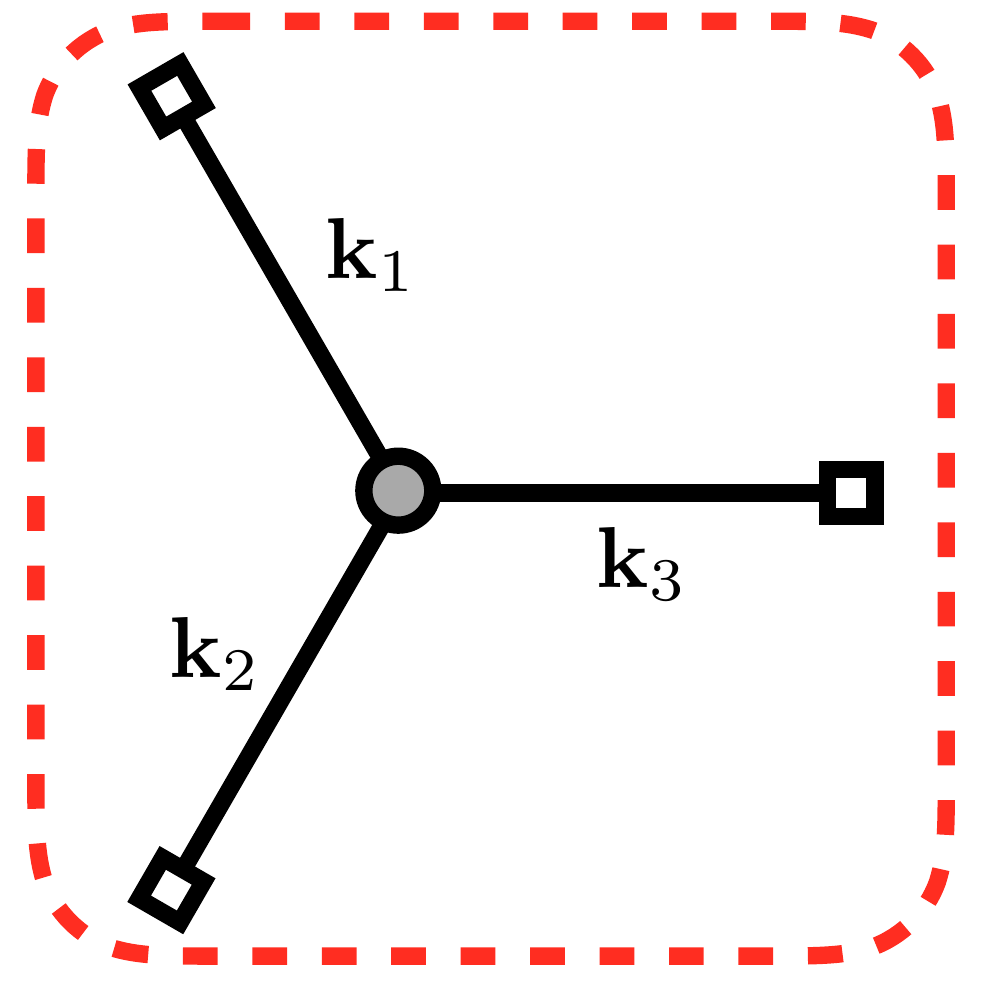}  
	\caption{The tree level diagram for three external legs.}
	\label{3ptexternaldiagram}
\end{figure}

We first consider the case with all the legs attached to the vertex being external. This corresponds to the tree level bispectrum
\begin{equation}
\begin{aligned}
\langle\zeta_{{\bf k}_1}\zeta_{{\bf k}_2}\zeta_{{\bf k}_3}\rangle'_{\cal D} &= \sum_{\cal A}\int_{-\infty}^0 d\tau_1~\left(-\frac{1}{\tau_1}\right) \frac{\partial}{\partial \tau_1} F_{a_1}(k_1,\tau_1) \frac{\partial}{\partial \tau_1} F_{a_1}(k_2,\tau_1) \frac{\partial}{\partial \tau_1} F_{a_1}(k_3,\tau_1)\\
&= \sum_{\cal A}\int_{-\infty}^0 d\tau_1~ k_1^2 k_2^2 k_3^2 (-\tau_1^2) G_{a_1}(k_1,\tau_1) G_{a_1}(k_2,\tau_1) G_{a_1}(k_3,\tau_1)\\
&= \left( k_1^2 k_2^2 k_3^2 \frac{\partial^2}{\partial K^2} \right) \sum_{\cal A}\int_{-\infty}^0 d\tau_1~ G_{a_1}(k_1,\tau_1) G_{a_1}(k_2,\tau_1) G_{a_1}(k_3,\tau_1)\\
&= \left( k_1^2 k_2^2 k_3^2 \frac{\partial^2}{\partial K^2} \right) \langle \phi_{{\bf k}_1} \phi_{{\bf k}_2} \phi_{{\bf k}_3} \rangle'_{\cal D} ~.
\end{aligned}
\end{equation}
In the last line we are just replacing $-\tau_1^2$ with $\displaystyle \frac{\partial^2}{\partial K^2}$ and pulling all the factors that are independent of time out of the time integral, where $K$ is any one of the momenta attached to the vertex at time $\tau_1$. The corresponding Feynman diagram is shown in FIG.~\ref{3ptexternaldiagram}, with the red dashed line corresponding to the part of diagram we have used the operator to convert to dS. 

\subsubsection{Two external legs}
Consider a general diagram ${\cal D}$ with a $a\zeta'^3$ vertex, and two of the legs from this vertex being external and one leg connected to the other parts of ${\cal D}$
\begin{equation}
\begin{aligned}
\langle\zeta_{{\bf k}_1}\zeta_{{\bf k}_2}\cdots\rangle'_{\cal D} = \sum_{\cal A}\int_{-\infty}^0 d\tau_1\int dT~\left(-\frac{1}{\tau_1}\right) \frac{\partial}{\partial \tau_1} F_{a_1}(k_1,\tau_1) \frac{\partial}{\partial \tau_1} F_{a_1}(k_2,\tau_1) \frac{\partial}{\partial \tau_1} \left[ {\cal V}_2 F_{a_1a_2}(l_1,\tau_1 ; l_2,\tau_2) \right]\cdots~,
\end{aligned}
\end{equation}
where $\int dT$ denotes all other time integrals (including the scale factor $a(\tau_i)$ associated to that vertex), and ${\cal V}_2$ comes from the possible space or time derivative from the vertex at time $\tau_2$, so ${\cal V}_2$ could be $1$, $\partial_{\tau_2}$ or $\pm i{\bf l}_2$. And without loss of generality, we have assumed the external lines attached to this vertex are ${\bf k}_1$ and ${\bf k}_2$. The subscript ${\cal D}$ is used to distinguish the contribution of ${\cal D}$ from the full $n$-point function. 

After evaluating the time derivatives, we may write
\begin{equation}
\begin{aligned}
\langle\zeta_{{\bf k}_1}\zeta_{{\bf k}_2}\cdots\rangle'_{\cal D} &= \sum_{\cal A}\int_{-\infty}^0 d\tau_1\int dT~\left(-\frac{1}{\tau_1}\right) k_1^2 k_2^2 l_1^2 \tau_1^3 G_{a_1}(k_1,\tau_1) G_{a_1}(k_2,\tau_1) {\cal V}_2 \tilde{F}_{a_1a_2}(l_1,\tau_1 ; l_2,\tau_2)\cdots \\
&= \sum_{\cal A}\int_{-\infty}^0 d\tau_1\int dT ~ k_1^2 k_2^2 l_1^2 (-\tau_1^2) G_{a_1}(k_1,\tau_1) G_{a_1}(k_2,\tau_1) {\cal V}_2 \tilde{F}_{a_1a_2}(l_1,\tau_1 ; l_2,\tau_2)\cdots \\
&= \left( k_1^2 k_2^2 l_1^2 \frac{\partial^2}{\partial K^2} \right) \sum_{\cal A}\int_{-\infty}^0 d\tau_1\int dT ~ G_{a_1}(k_1,\tau_1) G_{a_1}(k_2,\tau_1) {\cal V}_2 \tilde{F}_{a_1a_2}(l_1,\tau_1 ; l_2,\tau_2)\cdots ~,
\end{aligned}
\end{equation}
where $K$ is any one of the momenta attached to the vertex at time $\tau_1$. This equation can be understood as we have converted the part associated with the $\tau_1$ vertex to Minkowski space.
\begin{figure}[htbp]
	\centering    
	\includegraphics[scale=.5]{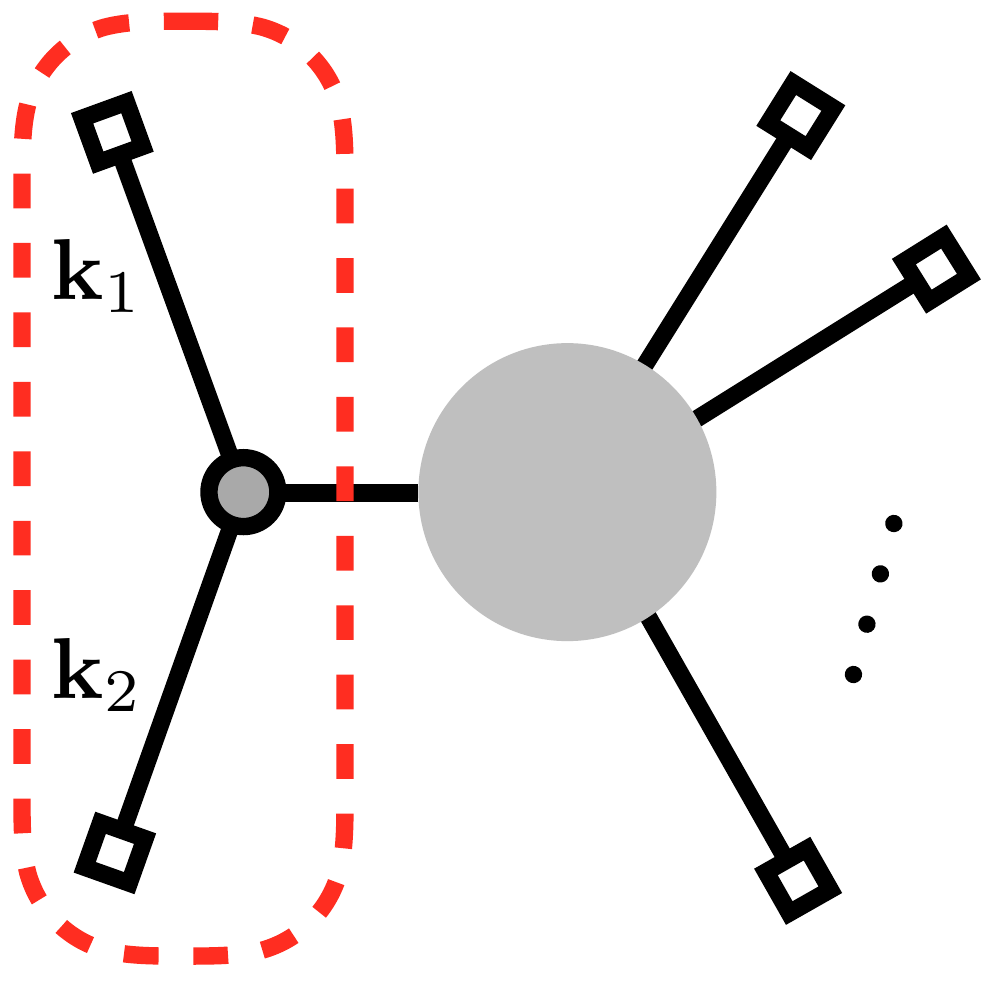}
	\caption{The tree level diagram for two external legs.}
	\label{2ptexternal}
\end{figure}

\subsubsection{One external legs}
Similarly, if only one leg is external, 
\begin{equation}
\begin{aligned}
\langle\zeta_{{\bf k}_1}\zeta_{{\bf k}_2}\cdots\rangle'_{\cal D} &= \sum_{\cal A}\int_{-\infty}^0 d\tau_1\int dT~\left(-\frac{1}{\tau_1}\right) \frac{\partial}{\partial \tau_1} F_{a_1}(k_1,\tau_1) \frac{\partial}{\partial \tau_1} {\cal V}_i F_{a_1a_i}(l_1,\tau_1 ; l_i,\tau_i) \frac{\partial}{\partial \tau_1} {\cal V}_j F_{a_1a_j}(l'_1,\tau_1 ; l_j,\tau_j)\cdots \\
&= \sum_{\cal A}\int_{-\infty}^0 d\tau_1\int dT~ k_1^2 l_1^2 l_1'^2 (-\tau_1^2) G_{a_1}(k_1,\tau_1) {\cal V}_i \tilde{F}_{a_1a_i}(l_1,\tau_1 ; l_i,\tau_i) {\cal V}_j \tilde{F}_{a_1a_j}(l'_1,\tau_1 ; l_j,\tau_j)\cdots \\
&= \left( k_1^2 l_1^2 l_1'^2 \frac{\partial^2}{\partial K^2} \right) \sum_{\cal A}\int_{-\infty}^0 d\tau_1\int dT ~ G_{a_1}(k_1,\tau_1) {\cal V}_i \tilde{F}_{a_1a_i}(l_1,\tau_1 ; l_i,\tau_i) {\cal V}_j \tilde{F}_{a_1a_j}(l'_1,\tau_1 ; l_j,\tau_j)\cdots \\
\end{aligned}
\end{equation}
where $l_i$ and $l_j$ are the internal momenta associate to the $i$-th and the $j$-th vertices, respectively.
\begin{figure}[htbp]
	\centering
	\includegraphics[scale=.5]{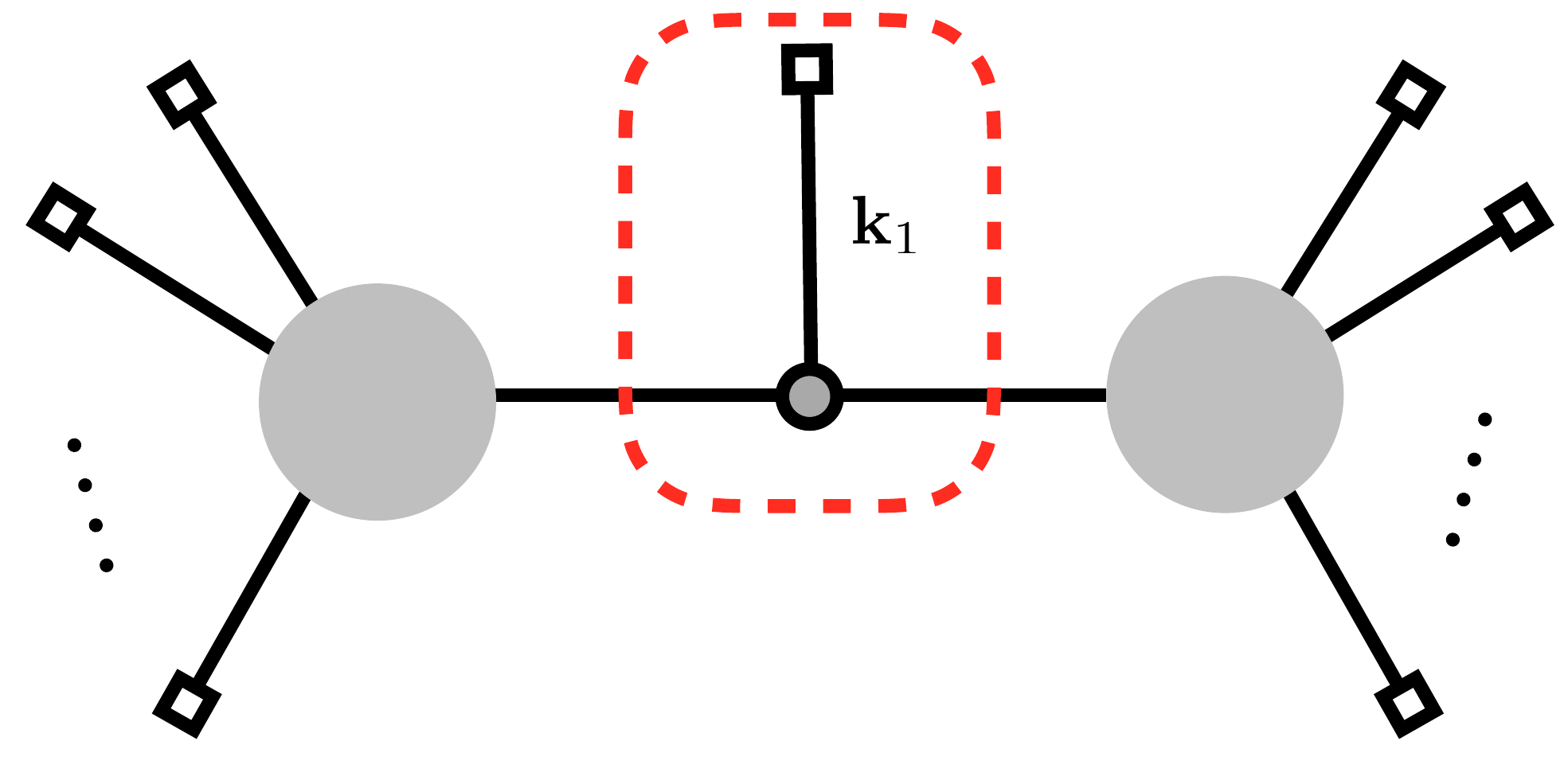}  
	\caption{The tree level diagram for one external leg.}
	\label{1ptexternal}
\end{figure}

\subsubsection{No external legs}
If there is no external legs
\begin{equation}
\begin{aligned}
\langle\zeta_{{\bf k}_1}\zeta_{{\bf k}_2}\cdots\rangle'_{\cal D} &= \sum_{\cal A}\int_{-\infty}^0 d\tau_1\int dT~\left(-\frac{1}{\tau_1}\right) \frac{\partial}{\partial \tau_1} {\cal V}_i F_{a_1a_2}(l_1,\tau_1 ; l_i,\tau_i) \frac{\partial}{\partial \tau_1} {\cal V}_j F_{a_1a_2}(l'_1,\tau_1 ; l_j,\tau_j) \frac{\partial}{\partial \tau_1} {\cal V}_k F_{a_1a_2}(l_1'',\tau_1 ; l_k,\tau_k) \cdots \\
&= \sum_{\cal A}\int_{-\infty}^0 d\tau_1\int dT~ l_1^2 l_1'^2 l_1''^2 (-\tau_1^2) {\cal V}_i \tilde{F}_{a_1a_i}(l_1,\tau_1 ; l_i,\tau_i) {\cal V}_j \tilde{F}_{a_1a_j}(l_1,\tau_1 ; l_j,\tau_j) {\cal V}_k \tilde{F}_{a_1a_k}(l'_1,\tau_1 ; l'_k,\tau'_k)\cdots \\
&= \left( l_1^2 l_1'^2 l_1''^2 \frac{\partial^2}{\partial K^2} \right) \sum_{\cal A}\int_{-\infty}^0 d\tau_1\int dT ~ {\cal V}_i \tilde{F}_{a_1a_i}(l_1,\tau_1 ; l_i,\tau_i) {\cal V}_j \tilde{F}_{a_1a_j}(l_1,\tau_1 ; l_j,\tau_j) {\cal V}_k \tilde{F}_{a_1a_k}(l'_1,\tau_1 ; l'_k,\tau'_k)\cdots ~,
\end{aligned}
\end{equation}
where $l_i$, $l_j$ and $l_k$ are the internal momenta associate to the $i$-th, $j$-th and $k$-th vertices, respectively. As mentioned before, the second derivative with respect to time will actually produce a contact term. So if any one of the ${\cal V}_2$, ${\cal V}_2'$ and ${\cal V}_2''$ is a time derivative, there will be a contact term. We will take care of it later.

Now we can conclude that a $a\zeta'^3$ vertex can be related to a $\phi^3$ vertex in the flat spacetime by an operator of the form, 
\begin{equation}
\boxed {a\zeta'^3 :~ k_1^2 k_2^2 k_3^2 \frac{\partial^2}{\partial K^2}~,}
\end{equation}
where $k_1$, $k_2$, $k_3$ are the momenta attached to the vertex and $K$ is any one of the $k_i$'s. 
\begin{figure}[htbp]
	\centering
	\includegraphics[scale=.4]{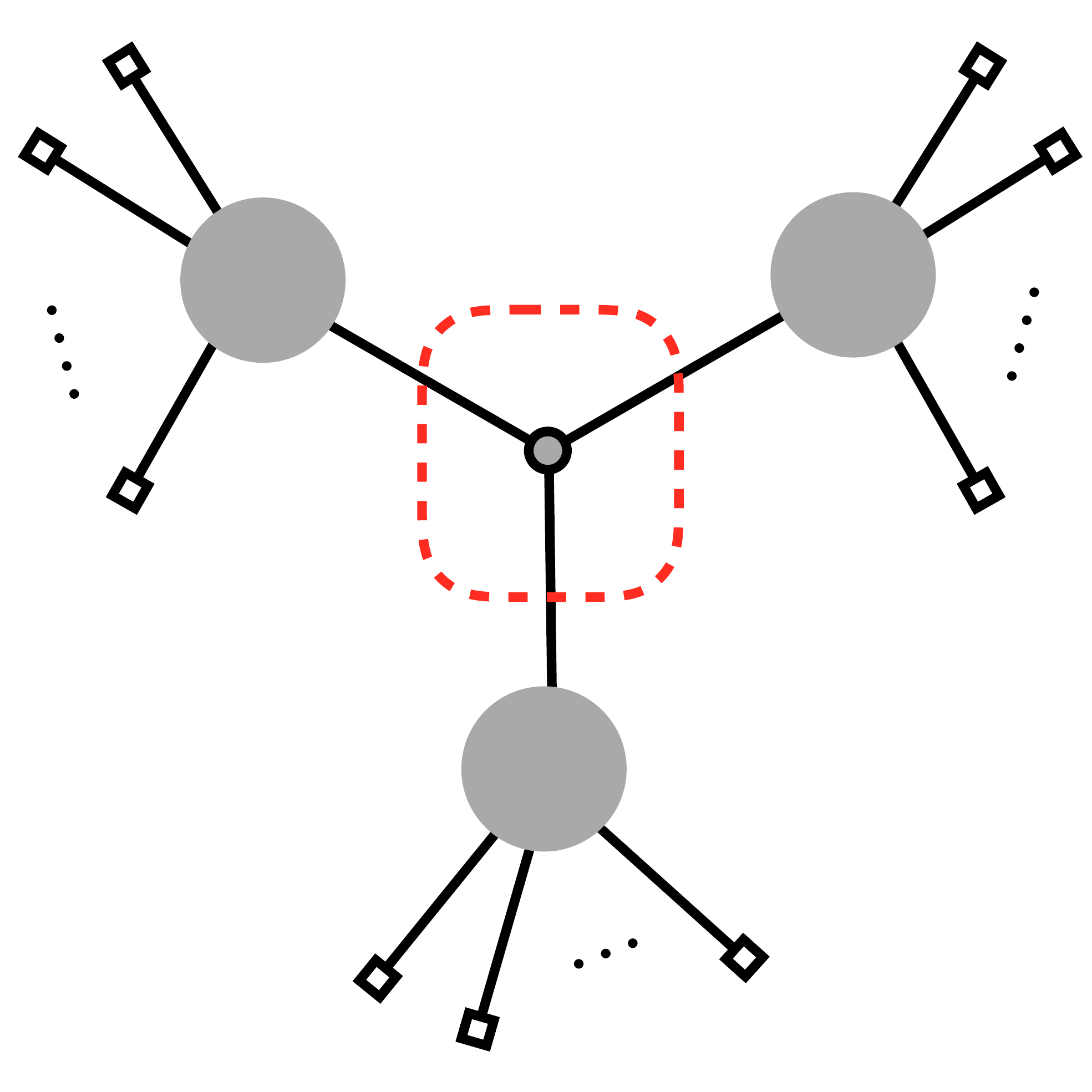}  
	\caption{The tree level diagram for no external leg.}
	\label{0ptexternal}
\end{figure}

\subsubsection{Contact term}
We will consider the case with two external legs attached to the $\tau_1$ vertex first. If the internal leg attached to this vertex is $\frac{\partial}{\partial \tau_1} \partial_2 F_{a_1a_2}(l_{12},\tau_1 ; l_{12},\tau_2)$, then the $a_1=a_2$ terms contain delta functions $\pm 2il_{12}^3\delta(\tau_1-\tau_2)$ and it act as a new interaction term $2\lambda^2 a^2 \zeta'^4$:
\begin{equation}
\begin{aligned}
(-i \lambda)^2 \int_{-\infty}^0 d\tau_1 \int_{-\infty}^0 d\tau_2 \int dT  \left(\frac{1}{\tau_1 \tau_2}\right) \bigg[ & \frac{\partial}{\partial \tau_1} F_+(k_1,\tau_1)  \frac{\partial}{\partial \tau_1} F_+(k_2,\tau_1) \left[2il_{12}^3\delta(\tau_1-\tau_2)\right] \partial_2 {\cal V}_i F_{++}(l_2,\tau_2 ; l_i,\tau_i) \partial_2 {\cal V}_j F_{++}(l_2,\tau_2 ; l_j,\tau_j) \cdots\\
+ & \frac{\partial}{\partial \tau_1} F_-(k_1,\tau_1) \frac{\partial}{\partial \tau_1} F_-(k_2,\tau_1) \left[-2il_{12}^3\delta(\tau_1-\tau_2)\right] \partial_2 {\cal V}_i F_{--}(l_2,\tau_2 ; l_i,\tau_i) \partial_2 {\cal V}_j F_{--}(l_2,\tau_2 ; l_j,\tau_j) \cdots \bigg]\\
= -2i \lambda^2 l_{12}^3 \left(- k_1^2 k_2^2 l_2^2 l_2'^2 \frac{\partial^2}{\partial K^2}\right) \int_{-\infty}^0 d\tau_1 \int dT  \bigg[ & G_+(k_1,\tau_1) G_+(k_2,\tau_1) {\cal V}_i \tilde{F}_{++}(l_2,\tau_1 ; l_i,\tau_i) {\cal V}_j \tilde{F}_{++}(l_2',\tau_1 ; l_j,\tau_j) \cdots\\
- & G_-(k_1,\tau_1) G_-(k_2,\tau_1) {\cal V}_i G_{--}(l_2,\tau_1 ; l_i,\tau_i) {\cal V}_j G_{--}(l_2,\tau_1 ; l_j,\tau_j) \cdots \bigg] ~.
\end{aligned}
\end{equation}
So for the $2\lambda^2 a^2 \zeta'^4$ vertex induced by $a\zeta'^3$ can be related to a $\phi^4$ vertex in the flat spacetime by
\begin{equation}
\boxed{ 2\lambda^2 a^2 \zeta'^4 :~ k_1^2 k_2^2 k_3^2 k_4^2\frac{\partial^2}{\partial K^2}~,}
\end{equation}
where $k_1$, $k_2$, $k_3$, $k_4$ are the momenta attached to the vertex and $K$ is any one of the $k_i$'s.

\subsection{$\displaystyle \frac{\mu}{2} a\zeta'(\partial\zeta)^2$ vertex}
Now we consider a general diagram with a $a\zeta'(\partial\zeta)^2$ vertex. The cases with different numbers of external legs can be done similarly as the $a\zeta'^3$ case.  What is different from the $a\zeta'(\partial\zeta)^2$ case is that the field associated to the internal line can be either $\zeta'$ or $\partial\zeta$. We will check that for these two cases, the differential operators extracted are of the same form. We will check with two of the legs from this vertex being external.

\subsubsection{The field associated to the internal line is $\zeta'$}
In this case we have
\begin{equation}
\begin{aligned}
\langle\zeta_{{\bf k}_1}\zeta_{{\bf k}_2}\cdots\rangle'_{\cal D} &= \sum_{\cal A}\int_{-\infty}^0 d\tau_1\int dT~\left(-\frac{1}{\tau_1}\right) (-{\bf k}_1\cdot{\bf k}_2) F_{a_1}(k_1,\tau_1)  F_{a_1}(k_2,\tau_1) \frac{\partial}{\partial \tau_1} {\cal V}_2 F_{a_1a_2}(l_1,\tau_1 ; l_2,\tau_2)\cdots \\ 
&= \sum_{\cal A}\int_{-\infty}^0 d\tau_1\int dT ~ ({\bf k}_1\cdot{\bf k}_2) l_1^2 \left[\left(1-k_1\frac{\partial}{\partial k_1}\right) G_{a_1}(k_1,\tau_1) \right] \left[\left(1-k_2\frac{\partial}{\partial k_2}\right) G_{a_1}(k_2,\tau_1) \right] {\cal V}_2 \tilde{F}_{a_1a_2}(l_1,\tau_1 ; l_2,\tau_2)\cdots \\
&= \left[ ({\bf k}_1\cdot{\bf k}_2) l_1^2 \left(1-k_1\frac{\partial}{\partial k_1}\right) \left(1-k_2\frac{\partial}{\partial k_2}\right) \right] \sum_{\cal A}\int_{-\infty}^0 d\tau_1\int dT ~ G_{a_1}(k_1,\tau_1) G_{a_1}(k_2,\tau_1) {\cal V}_2 \tilde{F}_{a_1a_2}(l_1,\tau_1 ; l_2,\tau_2)\cdots 
\end{aligned}
\end{equation}

\subsubsection{The field associated to the internal line is $\partial\zeta$}
In this case we have 
\begin{equation}
\begin{aligned}
\langle\zeta_{{\bf k}_1}\zeta_{{\bf k}_2}\cdots\rangle'_{\cal D} &= \sum_{\cal A}\int_{-\infty}^0 d\tau_1\int dT~\left(-\frac{1}{\tau_1}\right) (-{\bf k}_1\cdot{\bf l}_1) F_{a_1}(k_1,\tau_1) \frac{\partial}{\partial \tau_1} F_{a_1}(k_2,\tau_1) {\cal V}_2 F_{a_1a_2}(l_1,\tau_1 ; l_2,\tau_2)\cdots \\ 
&= \sum_{\cal A}\int_{-\infty}^0 d\tau_1\int dT ~ \left(-\frac{1}{\tau_1}\right) (-{\bf k}_1\cdot{\bf l}_1) k_2^2 \tau_1 \left[\left(1-k_1\frac{\partial}{\partial k_1}\right) G_{a_1}(k_1,\tau_1) \right] G_{a_1}(k_2,\tau_1) \left(1-l_1\frac{\partial}{\partial l_1}\right) {\cal V}_2 \tilde{F}_{a_1a_2}(l_1,\tau_1 ; l_2,\tau_2)\cdots \\
&= \left[ ({\bf k}_1\cdot{\bf l}_1) k_2^2 \left(1-k_1\frac{\partial}{\partial k_1}\right) \left(1-l_1\frac{\partial}{\partial l_1}\right) \right]  \sum_{\cal A}\int_{-\infty}^0 d\tau_1\int dT ~ G_{a_1}(k_1,\tau_1) G_{a_1}(k_2,\tau_1) {\cal V}_2 \tilde{F}_{a_1a_2}(l_1,\tau_1 ; l_2,\tau_2)\cdots 
\end{aligned}
\end{equation}
We can see that the operator has the same form as the previous case, the only difference is ${\bf k}_2 \leftrightarrow {\bf l}_1$. 

So, $a\zeta'(\partial\zeta)^2$ vertices can be related to $\phi^3$ vertices in the flat spacetime by an operator of the form
\begin{equation}
\boxed{a\zeta'(\partial\zeta)^2 :~ ({\bf k}_1\cdot{\bf k}_2) k_3^2 \left(1-k_1\frac{\partial}{\partial k_1}\right) \left(1-k_2\frac{\partial}{\partial k_2}\right) ~,}
\end{equation}
where $k_1$ and $k_2$ are the momenta correspond to the $\partial\zeta$ legs, and $k_3$ is the momentum corresponds to the $\zeta'$ leg.

\end{document}